\documentclass[journal]{IEEEtran}
\usepackage{authblk}
\usepackage{graphicx}
\usepackage{subcaption}
\usepackage{url}
\usepackage{array}
\usepackage[utf8]{inputenc} 
\usepackage[T1]{fontenc}
\usepackage{amsmath}
\usepackage{mathtools}
\usepackage{blkarray}
\usepackage{amsfonts}
\usepackage{amssymb}
\usepackage{amsthm}
\usepackage{enumerate}
\usepackage{cite}
\usepackage{lipsum}
\usepackage{mathtools}
\usepackage{cuted}
\usepackage{calc}
\usepackage{color}
\usepackage{epsfig}
\usepackage{setspace}
\usepackage{pstricks}
\usepackage{cancel}
\usepackage{multirow}
\usepackage{mathrsfs}
\usepackage{algorithm}
\usepackage{algpseudocode}
\usepackage{multirow}
\usepackage{diagbox}
\usepackage{multicol}
\usepackage{booktabs,makecell}
\usepackage{mathrsfs}

\usepackage{enumitem}

\usepackage{enumitem}
\newtheorem{defn}{Definition}
\newtheorem{thm}{{\cal T}heorem}

\newtheorem{constr}{Construction}
\newtheorem{remark}{Remark}

\newtheorem{example}{Example}

\newcommand{\W}{\mathcal{W}}

\newcommand{\Z}{\mathbb{Z}}

\newcommand{\F}{\mathbb{F}}
\newcommand{\M}{\mathcal{M}}

\newcommand{\U}{\mathcal{U}}

\newcommand{\J}{\mathcal{J}}

\input{epsf.sty}

\begin{document}
	\title{Novel Constructions for Computation and Communication Trade-offs in Private Coded Distributed Computing}
		\author{
		\IEEEauthorblockN{Shanuja Sasi, ~\IEEEmembership{Member, ~IEEE} and
			Onur Günlü, ~\IEEEmembership{Senior Member, ~IEEE} }
			\thanks{This work was partially supported by the ZENITH Research and Leadership Career Development Fund under Grant ID23.01, and the Swedish Foundation for Strategic Research (SSF) under Grant ID24-0087.  This article was presented in part at the IEEE Int. Symp. Inf. Theory (ISIT 2025) \cite{con}.}
		\thanks{S. Sasi is with the Department of Electrical Engineering, Indian Institute of Technology, Kanpur - 208016  (e-mail: shanujas@iitk.ac.in). O. Günlü is with the Lehrstuhl für Nachrichtentechnik, TU Dortmund, Germany (email: onur.guenlue@tu-dortmund.de) and with the Information Theory and Security Laboratory (ITSL), Linköping University, SE-581 83 Linköping, Sweden. }
	}
	\maketitle
	\begin{abstract}
		Distributed computing enables scalable machine learning by distributing tasks across multiple nodes, but ensuring privacy in such systems remains a challenge. This paper introduces a novel \textit{private coded distributed computing}  model that integrates privacy constraints to keep task assignments hidden. By leveraging \textit{placement delivery arrays} (PDAs), we design an extended PDA framework to characterize achievable computation and communication loads under privacy constraints. By constructing two classes of extended PDAs, we explore the trade-offs between computation and communication, showing that although privacy increases communication overhead, it can be significantly alleviated through optimized PDA-based coded strategies.
	\end{abstract}
\section{Introduction}
\textit{Distributed computing (DC)} methods play a crucial role in enabling scalable machine learning, especially when dealing with large datasets. These methods decompose complex processes into smaller, parallelizable tasks that can be distributed across multiple computing nodes. For example, in large-scale text classification tasks, a distributed framework can efficiently extract features, such as word frequencies, from subsets of documents assigned to different nodes \cite{Mapreduce}. These features are subsequently reorganized and transmitted to designated nodes for further processing, such as aggregating results or updating model parameters in algorithms like logistic regression. This structured workflow ensures efficient utilization of computational resources and seamless management of large-scale data. Several frameworks demonstrate the practical application of these principles. For instance, Apache Spark \cite{apache} leverages its distributed datasets to parallelize data processing and includes the MLlib library for implementing distributed machine learning algorithms. 

In this paper, we consider MapReduce framework \cite{Mapreduce} where the computing task can be divided into three major phases. Initially, a \textit{Map phase} is responsible for tasks such as feature extraction and transformation. During this phase, each node processes a subset of the data, extracting relevant features and producing intermediate outputs. Following this, a \textit{Shuffle phase} organizes these intermediate results based on specific criteria, such as feature indices or class labels. This step ensures that related data is efficiently routed to nodes tasked with subsequent operations. Finally, a \textit{Reduce phase} consolidates the processed outputs, combining intermediate results to compute final output functions, such as gradients or model predictions. 

Coding-theoretic techniques have been extensively applied in DC for a wide range of applications, including  coded matrix multiplication \cite{MCJ}, gradient computations \cite{OGU,TLDK,SAR}, and coded MapReduce \cite{LMA,YYW,SFZ,WCJ,DZZWL,PLSSM,WCJnew,SGR,SG,SG2,SG3}. Some of these works leverage {\it placement delivery arrays (PDAs)} \cite{JQ,YTC}. 
Initially introduced in \cite{YCTCPDA} as a solution for coded caching problems, PDAs have since become a widely used design tool for addressing various challenges. 
Some works~\cite{Kumar2014ComprehensiveReviewStraggler} examined the challenges posed by straggler nodes, others identified the data shuffling as a critical bottleneck in distributed applications. For example, the study in~\cite{Zhuoyao2013PerformanceModelingMapReduce}, which analyzed the behavior of various algorithms on the Amazon EC2 cluster, found that communication overhead during the data shuffling phase was the primary constraint in performing distributed tasks. 
In \cite{LMA}, coding techniques are leveraged to achieve a reduction in communication load, proportional to the computation load, compared to uncoded schemes. This was accomplished by replicating files across servers to create coding opportunities during the data shuffling phase.  
Since its introduction in~\cite{LMA}, this coded  framework has been extended to accommodate various scenarios. 
For example, in heterogeneous environments where nodes are assigned different numbers of files for mapping and different numbers of functions for computation,  novel schemes based on hypercube and hypercuboid geometries were proposed in~\cite{Woolsey2021NewCombinatorialCoded, Woolsey2021CombinatorialDesignCascaded}. The original scheme in~\cite{LMA}, despite being information-theoretically optimal, requires an exponentially growing number of input files with the number of computing nodes, which can be impractical. 
Other studies explored DC with different communication paradigms. For instance, DC over wireless channels was investigated in~\cite{Li2017ScalableFrameworkWireless}, and scenarios with storage and computational resource constraints at nodes were analyzed in~\cite{YYW}. 
 In \cite{YYW}, the authors adopted PDA designs to develop a coding framework, focusing on characterizing the storage-computation-communication trade-off rather than the computation-communication trade-off explored in \cite{LMA}. In \cite{onur}, the authors developed a framework for secure multi-function computation, characterizing tight rate regions for both lossless and lossy settings while addressing secrecy and privacy constraints using a remote source model. In addition to communication bottlenecks, several other key areas have been explored to enhance DC. To address delays caused by slow workers in distributed systems, several methods using codes have been introduced to speed up DC  \cite{2,4}. The importance of low-latency operations in distributed systems is emphasized by \cite{5}, which discusses timely coded computing as a means to reduce overall system delays.
 
 Our work focuses on protecting the indices of computed functions, motivated by practical scenarios where task allocation patterns themselves are sensitive.
 	 For instance, in federated learning \cite{N1}, while gradient functions  are public, revealing which model parameter a node is updating could leak insights into data distributions or model architecture. Similarly, in multi-tenant cloud systems \cite{N2}, masking function indices prevents users from inferring neighboring tasks (e.g., financial modeling vs. climate simulations), which might expose operational priorities or vulnerabilities.
 	
 	Unlike \textit{secure multi-party computation} (SMPC) \cite{R1} or homomorphic encryption methods \cite{R2}, designed to protect the content of data, our approach obscures task-node relationships without requiring data encryption.
 	In contrast to differential privacy \cite{R3,RO}, which distorts outputs to prevent leakage, our coded scheme avoids accuracy degradation by leveraging redundancy in computations.
 	Similarly, while collusion-resistant schemes (e.g., secret-sharing \cite{R4}) defend against adversarial node alliances, our model emphasizes efficiency by operating under a non-colluding node assumption.
 	Though SMPC offers strong guarantees, it incurs high overhead, whereas we focus on schemes based on the PDA framework, which provides lightweight, information-theoretic task privacy, bridging the gap between cryptographic methods and non-private coded computing.
 	
 	This work advances the use of PDAs beyond their traditional role in coded caching \cite{R5} by adapting them to the privacy-preserving \textit{coded distributed computing} (CDC) setting. While PDAs in caching optimize multicast opportunities and bandwidth usage, their role in privacy-preserving CDC is redefined to ensure task assignment privacy. 
 	This paper positions its contribution within the broader literature on demand-private coded caching and \textit{private information retrieval} (PIR). Prior work by Wan et al. \cite{R6} formalized demand privacy constraints, proposing a virtual user strategy where each real user is paired with multiple virtual users to ensure uniform file requests. This symmetry prevents users from distinguishing real demands, achieving perfect privacy. However, the scheme's subpacketization grows exponentially with the number of users, rendering it impractical. To address this, in \cite{R6}, another MDS-coded scheme was developed, offering significantly lower subpacketization while maintaining order-optimal rates. 
 	Extending beyond demand privacy, the problem of securing both demands and caches requires preserving the confidentiality of user requests and cached content. This dual privacy constraint, studied in \cite{R8}, ensures no leakage about user preferences or stored data. PIR \cite{R9} enables a user to privately retrieve a file from distributed servers without revealing which file is requested. Advanced PIR schemes ensure privacy even if multiple servers collude or store coded data. However, both demand-private coded caching and PIR focus on a centralized model where users interact with a centralized server. Crucially, PIR and demand-private coded caching differ from our problem, as while both of them relies on a centralized server for transmissions, our setup involves nodes cooperatively exchanging data to compute output functions. The proposed framework instead shares more in common with decentralized coded caching schemes, such as those in \cite{R10}, but uniquely applies PDAs to manage computation privacy.
 	To achieve this, we introduce a novel approach where two PDAs are combined using a block structure involving virtual nodes. Each physical (real) node randomly selects a column from this structure, masking its true role. This contrasts sharply with classical PDAs, which map users deterministically without privacy concerns. During the Shuffle phase of computation, the intermediate results are split and XORed across special partitions. This encoding ensures that the transmitted symbols do not leak any information about the underlying function indices.
 	This work theoretically quantifies the communication overhead introduced by privacy requirements. Theorems 2 and 3 in this paper demonstrate how ensuring function-index privacy increases the communication load and how optimized PDA designs can mitigate this penalty.

 	Overall, our key contribution is a privacy-preserving, decentralized coded computing model that leverages PDAs not only for content delivery but to obscure task-node assignments. This repurposing of combinatorial tools to private distributed computations without cryptographic burdens marks a significant paradigm shift, blending ideas from coded caching, PIR, and distributed computing into a novel framework for practical, privacy-sensitive scenarios.
\subsection{Main Contributions}
The main contributions in this paper are summarized as follows.
\begin{itemize}
	\item We integrate a privacy constraint into DC to ensure that the task assigned to each node (i.e., the output function index) remains hidden from other nodes. This guarantees that no node can infer another’s specific role or data characteristics, thereby preserving privacy. We refer to DC models with such privacy constraints as \textit{private coded distributed computing (private CDC) models}.  
	
	\item First, we construct an extended PDA from two given PDAs and develop a private CDC scheme based on this extended structure.
	\item  We introduce two sets of extended PDAs and use them to characterize the achievable \textit{computation-communication trade-offs} for private CDC models.  
	
	\item From the first set of extended PDAs, we demonstrate that the computation load achievable in private CDC models is {identical} to that in non-private CDC models. However, the communication load in the private setting is {scaled by a factor of} \( Q \), which is the number of output functions, compared to the non-private scenario.  
	
	\item To optimize this trade-off, we construct a second set of extended PDAs that {allow a reduction in communication load}. In this case, the computation load in private CDC models {increases by an additive factor} compared to the non-private setting, while the communication load is {scaled by a factor smaller than} \( Q \).
\end{itemize}

\noindent
{\it Organization of this paper:} Section \ref{problem defintion} provides an overview of DC frameworks and PDAs. In Section \ref{dlmodel}, we introduce the concept of private CDC models. Our main results are presented in Section \ref{main}, where we detail the extended PDA construction and the coded scheme designed for private CDC models. 
Additionally, we present the construction of the first set of extended PDAs in Section \ref{trade}, where we also characterize the achievable trade-offs between computation and communication loads. In Section \ref{opt}, we introduce another set of extended PDAs to derive improved computation-communication trade-offs.

\noindent{\it Notation:}  The bit wise exclusive OR (XOR) operation is denoted by $\oplus.$ The notation $[n]$ represents the set $\{1,2, \ldots , n\}$.
For any $m \times n$ array  $\textbf{A} = [a_{i,j}],$ for $ i \in [m] $ and $j \in [n],$  the array $\textbf{A}+b$, is defined as $\textbf{A}+b =[a_{i,j}+b]$. For any given set of integers $A, t$-subset of the set $A$ refers to the subsets of size $t$ from the set $A$, i.e. $\{U: U \subseteq A, |U| = t\}$.
	\section{Background and Preliminaries}
	\label{problem defintion}
	\subsection{DC Background}
	
We examine a DC model, as discussed in \cite{LMA}, which consists of \( K \) nodes labeled from \( [K] \). Each node is responsible for computing one of \( Q \) possible output functions, denoted as \( \{\phi_q : q \in [Q]\} \). Each output function \( \phi_q \), where \( q \in [Q] \), processes a collection of \( N \) input data files, each containing \( w \) bits, and produces a stream of \( b \) bits as output.
The output function $\phi_q$, for $q \in [Q]$, is defined as:
\begin{align}
	\phi_q : \mathbb{F}_{2^w}^N \to \mathbb{F}_{2^b}
\end{align}
where it maps $N$ input files $\mathcal{W} = \{W_1, \ldots, W_N\}$, each with $w$-bit contents, to a single output of $b$-bit value.
The computation of $\phi_q$ is broken into the following two stages.
\begin{itemize}
	\item A map function $g_q$:
	\begin{align}
		g_q : \mathbb{F}_{2^w} \to \mathbb{F}_{2^{\alpha}}
	\end{align}
	which maps each input file $W_n$ into an {\it intermediate value (IV)} $v_{q,n} = g_{q}(W_n) \in \F_{2^{\alpha}}$.    
	\item A reduce function $h_q$:
	\begin{align}
		h_q : \mathbb{F}_{2^{\alpha}}^N \to \mathbb{F}_{2^b}
	\end{align}
	which aggregates the  IVs $\{v_{q,n}\}_{n=1}^N$ into the final $b$-bit output.
\end{itemize}
Combining the intermediate mapping and reduce functions, the output function $\phi_q$ can be expressed as
\begin{align}
	\label{out}
	\phi_q(\W) = h_q(v_{q,1}, \ldots , v_{q,N}  ) 
	= h_q(g_{q}(W_1), \ldots , g_{q}(W_{N}) ).
\end{align}
	\noindent
	The process of function computation in this DC framework occurs in three main phases:
	\begin{enumerate}[leftmargin=*,label=\arabic*.]
		\item \textbf{Map Phase:} 
		Each node $k \in [K]$ stores a subset of the input files $\mathcal{M}_k \subseteq \mathcal{W}$ and computes the IVs
		\begin{align}
		\{v_{q,n} : q \in [Q], W_n \in \mathcal{M}_k, n \in [N]\}.
		\end{align}
		\item \textbf{Shuffle Phase:} 
		Each node $k \in [K]$ is responsible for computing an assigned output function $\phi_{d_k}$. To do so, it needs the IVs from files it does not store locally. The missing IVs for node $k$ are denoted as
		\begin{align}
		\{v_{{d_k},n} : W_n \in \mathcal{W} \setminus \mathcal{M}_k, n \in [N]\}.
		\end{align}
		To share the missing IVs of other nodes, each node $k$ generates a bit sequence $\mathbf{X}_k \in \{0, 1\}^{l_k}$ based on its local IVs and broadcasts it to all other nodes.
		
		\item \textbf{Reduce Phase:} 
		Upon receiving sequences $\{\mathbf{X}_j\}_{j \in [K] \setminus k}$ from other nodes, each node $k \in [K]$ decodes all necessary IVs to compute its output function $\phi_{d_k}$.
	\end{enumerate}
	\noindent
	We next define the computation and communication loads.
	\begin{defn}[Computation Load \cite{LMA}]
		Computation load $r$ is the total number of bits associated with the files mapped across $K$ nodes, normalized by the total size of the files. Mathematically, the computation load \( r \) is formally defined as
		\begin{equation}
			r = \frac{\sum_{k=1}^K |\mathcal{M}_k| \cdot w}{N \cdot w} = \frac{\sum_{k=1}^K |\mathcal{M}_k|}{N}
		\end{equation}
		where \( |\mathcal{M}_k| \) is the number of files stored at node \( k \), \( N \) is the total number of files, and \( w \) is the file size in bits.
	\end{defn}
	\begin{defn}[Communication Load \cite{LMA}]
	 Communication load $L$ is the total number of bits transmitted by the $K$ nodes during the Shuffle phase, normalized by the number of bits in all IVs. The communication load \( L \) is defined as
	\begin{equation}
		L = \frac{\sum_{k=1}^K \ell_k}{Q \cdot N \cdot \alpha}
	\end{equation}
	where \( \ell_k \) is the length of the transmitted sequence \( \mathbf{X}_k \), \( Q \) is the number of output functions, and \( \alpha \) is the size of each IV in bits.
	\end{defn}
		The communication load measures the efficiency of information exchange during the Shuffle phase. 
	The tradeoff between computation and communication loads is established for this setup in \cite{LMA}. The \textit{optimal} tradeoff curve is given by the lower convex envelope of $
	\left\{\left(r, L^*(r)\right) : r \in [K]\right\},
	$
	where we have
	\begin{align}
	\label{cdc}
	L^*(r) \triangleq \frac{1}{r}\left(1 - \frac{r}{K}\right).
	\end{align}
	\subsection{PDA Background}
	Yan et al. \cite{YCTCPDA} introduced the concept of PDA to represent coded caching schemes with the goal of reducing sub-packetization levels. Since then, several coded caching schemes based on the PDA concept have been reported.
	\begin{defn}  ({\bf PDA}\cite{YCTCPDA}):
		\label{def pda}
		For positive integers $K, F, Z,$ and $S,$ an $F \times K$ array $P = [p_{f,k}]$ with $ f \in [F],$ and $ k \in [K]$ composed of a specific symbol $*$ and $S$ positive integers $[S],$ is called a $(K, F, Z, S)$ placement delivery array (PDA) if it satisfies the following conditions:
		\begin{itemize}
			\item {\it A1:} The symbol $*$ appears $Z$ times in each column;
			\item {\it A2:} Each integer occurs at least once in the array;
			\item {\it A3:} For any two distinct entries $p_{f_1,k_1}$ and $p_{f_2,k_2}, s=p_{f_1,k_1} = p_{f_2,k_2} $ is an integer only if
			\begin{enumerate}
				\item $f_1$ $\neq f_2$ and $ k_1$ $\neq k_2,$ i.e., they lie in distinct rows and distinct columns; and
				\item $p_{f_1,k_2} = p_{f_2,k_1} = *,$ i.e., the corresponding $2 \times 2$ sub-array formed by rows $f_1, f_2$ and columns $k_1, k_2$ must be either of the following forms
				$ \begin{pmatrix}
				s & *\\
				* & s
				\end{pmatrix} $ or 
				$\begin{pmatrix}
				*& s\\
				s & *
				\end{pmatrix}.$\qed
			\end{enumerate} 
		\end{itemize}
		\label{def:PDA}
	\end{defn}
	\begin{example}
		\label{PDA example}
		Consider an $4 \times 5$ array $P_1$ as given below. It satisfies conditions \textit{A1, A2} and \textit{A3}. There are $2$ stars in each column and a total of $4$ integers in the array. Hence,  $P_1$ is a $(5,4,2,4)$ PDA
		{
		\begin{equation}
		\label{A1}
		P_1 =
		\begin{blockarray}{ccccc}
		\begin{block}{(ccccc)}
		* & * & * & 1 &  2 \\
		* & 1 & 2 & * & * \\
		1 & * & 3 & * &  4\\
		2 & 3 & * & 4 &  *\\
		\end{block}
		\end{blockarray}. 
		\end{equation}}
	\end{example}
		\begin{defn} ({\bf $g-$regular PDA}\cite{YCTCPDA}):
		An array $P$ is said to be a $g$-$(K, F, Z, S)$ PDA or $g$-regular $(K, F, Z, S)$ PDA if it satisfies A1, A3, and the following condition
		\begin{itemize}
			\item {\it $A2'$:}  Each integer appears $g$ times in $P$, where $g$ is a constant.\qed
		\end{itemize}
		\label{def:g-PDA}
	\end{defn}
	\begin{example}
		\label{g regular example}
		The $4 \times 6$ array $P_2$ provided below is a $3$-$(6,4,2,4)$ PDA
		\begin{equation}
			\label{A2}
			P_2 =
			\begin{blockarray}{cccccc}
				\begin{block}{(cccccc)}
					* & * & * & 1 &  2&3  \\
					* & 1 & 2 & * & *& 4 \\
					1 & * & 3 & *  &  4 &*\\
					2 & 3 & * & 4 & *  &*\\
				\end{block}
			\end{blockarray}.
		\end{equation}
	\end{example}	
	A PDA is a combinatorial tool that balances storage, computation, and communication in distributed systems. It is represented as a $K \times F$ matrix, where 
		$F$ denotes the number of data batches (subdivided input files) and 
		$K$ is the number of nodes. Each column corresponds to a node, where stars ($*$) denote locally stored batches (parameter $Z$), reducing communication needs, while integers represent coded messages (parameter $S$) exchanged to recover missing data during computation. For example, a node storing 
		$Z=2$  batches (marked by $*$) avoids transmitting raw data for those batches, whereas integers represent coded multicast messages that enable nodes to collaboratively reconstruct missing information.

	Suppose that we are given a $(K, F,Z,S)$ PDA $P= [p_{f,k}]$ for $ f\in [F],k\in[K]$, and  some integers $ K,F,Z,$ and $S$ such that each integer appears more than once in the PDA $P$. In \cite{YTC}, a coded  scheme for a DC model is derived from this PDA where there are  $K$  nodes, and $\eta F$ number of files, for some positive integer $\eta$. This DC model achieves a computation load $r=\frac{ZK}{F}$ and communication load 
\begin{align}
	\label{cdc pda comm load}
	L_{pda}= \frac{S}{KF}+\sum_{g=2}^{K}\frac{S_g}{KF (g-1)}= \sum_{g=2}^{K} \frac{g S_g}{KF (g-1)}
\end{align}	
where $S_g$ is the number of integers in $[S]$ which appear exactly $g$ times in the PDA $P$. This equation characterizes the communication load in terms of the parameters of the PDA.
Here, \(K\) denotes the total number of nodes, \(F\) is the number of file batches, \(S\) is the total number of distinct integers in the PDA, and \(S_g\) counts the integers that occur exactly \(g\) times in the PDA. The summation term captures the multicast opportunities that arise when certain integers occur multiple times. Specifically, an integer appearing \(g\) times allows a single coded transmission to satisfy the demands of \(g-1\) nodes. This saving is reflected in the denominator factor \((g-1)\), since one transmission can replace \(g-1\) individual transmissions. Meanwhile, for that same integer, \(g\) coded transmissions are sent by \(g\) different nodes, leading to the factor \(g\) in the numerator. This interpretation emphasizes how the structure of the PDA directly determines the communication efficiency. 
 \section{Private CDC Set-up}
 \label{dlmodel}
In the proposed private CDC model, which involves $K$ computational nodes indexed by $[K]$, the process of function computation is carried out in three primary phases:

\begin{enumerate}[leftmargin=*,label=\arabic*.]
	\item \textbf{Map Phase:} 
	Each node $k \in [K]$ generates a local random variable $a_k$, which is independent of the database $\mathcal{W}$ and independent across nodes. This randomness $a_k$ is only accessible to node $k$. Subsequently, the node stores a subset of the input files $\mathcal{M}_k \subseteq \mathcal{W}$ and computes the corresponding IVs, satisfying the condition
	\begin{align}
	H(\mathcal{M}_k \,|\, a_k, \mathcal{W}) = 0.
	\end{align}
	\item \textbf{Shuffle Phase:} 
	Each node $k \in [K]$ is assigned to compute a specific output function $\phi_{d_k}$. Given its randomness $a_k$, stored subset $\mathcal{M}_k$, and the assigned function $\phi_{d_k}$, the node $k$ generates and broadcasts a query $\mathbf{y}_k$ to all other nodes. 
	
	\textit{Privacy Constraint:} For each $k \in [K]$, it is required that no other node gains any knowledge about the index $d_k$ of the output function assigned to node $k$. Privacy is maintained if the following holds for every $k \in [K]$
	\begin{align}
	I\big(\{d_j\}_{j \in [K]}; \{\mathbf{y}_j\}_{j \in [K]} \,|\, d_k, \mathcal{M}_k\big) = 0.
	\end{align}
	After receiving all the queries, node $k$ transmits a coded symbol $\mathbf{X}_k$ to all other nodes such that
	\begin{align}
	H\big(\mathbf{X}_k \,|\, \mathcal{M}_k, a_k, \{\mathbf{y}_j\}_{j \in [K]}\big) = 0.
	\end{align}
	\item \textbf{Reduce Phase:} 
	Upon receiving the broadcast coded symbols $\{\mathbf{X}_j\}_{j \in [K] \setminus k}$ from the other nodes, each node $k \in [K]$ decodes the necessary information to compute its assigned output function $\phi_{d_k}$. This decoding is successful if we have
	\begin{align}
	H\big(\phi_{d_k} \,|\, \mathcal{M}_k, a_k, d_k, \{\mathbf{y}_j\}_{j \in [K]}, \{\mathbf{X}_j\}_{j \in [K]}\big) = 0.
	\end{align}
\end{enumerate}

	\section{Coding Scheme for Private CDC using PDA}
\label{main}
In this section, we generate an extended PDA from two given PDAs and provide a private CDC scheme using the extended PDA. Construction of an extended PDA ${\bf P}$ from two PDAs ${\bf P}^{(1)}$ and ${\bf P}^{(2)}$ is provided in {\bf Algorithm \ref{algo1}}. The proof of correctness of {\bf Algorithm \ref{algo1}} is provided in Appendix \ref{proof algo}. Moreover, in Theorem \ref{thm1}, using this extended PDA, a private CDC scheme is obtained, the proof of which is provided in Appendix \ref{proof thm1}.

\textbf{Algorithm \ref{algo1}} constructs an extended PDA $\mathbf{P}$ by merging two PDAs, $\mathbf{P}^{(1)}$ and $\mathbf{P}^{(2)}$. The inputs are, $\mathbf{P}^{(1)}$ which is a $(K_1, F_1, Z_1, S_1)$ PDA defining $K_1$ nodes, and  $\mathbf{P}^{(2)}$ which is  a $(K_2, F_2, Z_2, S_2)$ PDA introducing $Q = K_2$ output functions to mask task-node mappings. Hence, the PDA $\mathbf{P}$ is structured as a block array with $F_1$ row blocks (total batches: $F_1F_2$) and   $K_1$ column blocks (total nodes: $K_1K_2$). The system involves $K_1K_2$ effective nodes, where $K_1$ nodes are real nodes (physical machines) and $(K_2-1)K_1$ nodes are virtual nodes (abstract entities to mask task assignments).  Within each column block there is one column corresponding to the real node and the rest of the columns is to mask task assignments.  Each node stores $Z = Z_1F_2 + (F_1 - Z_1)Z_2$ batches, balancing computation and communication. Virtual nodes impersonate randomized roles to hide real nodes' tasks. This structure ensures real nodes compute tasks while virtual nodes mask their roles, achieving privacy through combinatorial redundancy. While the algorithm defines the replacement rules for integers and stars, it is not immediately obvious that the resulting array $\mathbf{P}$ satisfies the PDA properties (Definition~3). Appendix \ref{proof algo} formally proves that $\mathbf{P}$ is indeed a valid PDA with parameters $(K_1 K_2, F_1 F_2, Z_1 F_2 + (F_1 - Z_1) Z_2, S_1S_2)$.
	 \begin{algorithm}
	\caption{Construction of a $(K_1K_2,F_1F_2,Z_1F_2 + (F_1-Z_1)Z_2,S_1S_2)$ PDA  ${\bf P}$ extended from  $(K_1,F_1,Z_1,S_1)$   PDA ${\bf P}^{(1)}$, where each integer appears more than once and $(K_2,F_2,Z_2,S_2)$  PDA ${\bf P}^{(2)}$ where $K_2 >1$.}
	\label{algo1}
	${\bf P}$ is obtained by replacing the entries in ${\bf P}^{(1)}$ as follows:\begin{enumerate}
		\item Replace each integer $s\in [S_1]$ in ${\bf P}^{(1)}$ by the PDA \\${\bf P}^{(2)}+ (s-1)S_2$, where $* + (s-1)S_2 = *$.
		\item Replace each $*$ in ${\bf P}^{(1)}$ by a $F_2 \times K_2$ array denoted as $ {\bf X}$,  where all the entries are represented by the symbol $*$.
	\end{enumerate}
\end{algorithm}

\begin{example}
	\label{ex algo 1}
	This example illustrates {\bf Algorithm \ref{algo1}}.
		The two given arrays in  {\bf Algorithm \ref{algo1}} are defined as
	\begin{align}
		A^{(1)} =
	\begin{bmatrix}
		\ast & 1& \ast & 3 \\
		\ast &2&3 & \ast \\
		1 & \ast & \ast&4\\
		2&*&4&*
	\end{bmatrix} \text{ and } 	A^{(2)} =\begin{bmatrix}
	\ast & 1 \\
	1&*
	\end{bmatrix}.
	\end{align}
	The array $A^{(1)}$ is a $(4,4,2,4)$ PDA and $A^{(2)}$ is a  $(2,2,1,1)$ PDA.
		An extended array \( {\bf A}_1 \) is constructed from \( A^{(1)} \) and \( A^{(2)} \) as follows:
	\begin{itemize}
		\item Replace each integer \(1\) in \(A^{(1)}\) with \(A^{(2)}\), i.e., we have $\begin{bmatrix}
		\ast & 1 \\
		1&*
		\end{bmatrix}.$
		\item Replace each integer \(2\) in \(A^{(1)}\) with \(A^{(2)}+1\), i.e., we have $\begin{bmatrix}
		\ast & 2 \\
		2&*
		\end{bmatrix}.$
		\item Replace each integer \(3\) in \(A^{(1)}\) with \(A^{(2)}+2\), i.e., we have $\begin{bmatrix}
			\ast & 3 \\
			3&*
		\end{bmatrix}.$
		\item Replace each integer \(4\) in \(A^{(1)}\) with \(A^{(2)}+3\), i.e., we have $\begin{bmatrix}
		\ast & 4 \\
		4&*
		\end{bmatrix}.$
		\item Replace each \(\ast\) in \(A^{(1)}\) with a \(2 \times 2\) block of all \(\ast\)'s:
		$
		\begin{bmatrix}
			\ast & \ast \\
			\ast & \ast 
		\end{bmatrix}.
		$
	\end{itemize}
	The resulting extended array ${\bf A}_1$ is:
	\begin{align}
	{\bf A}_1 =
	\begin{bmatrix}
		\begin{bmatrix}
			\ast & \ast \\
			\ast & \ast 
		\end{bmatrix} &\begin{bmatrix}
		\ast & 1 \\
		1&*
		\end{bmatrix}&
		\begin{bmatrix}
			\ast & \ast \\
			\ast & \ast 
		\end{bmatrix} &
		\begin{bmatrix}
		\ast & 3 \\
		3&*
		\end{bmatrix} \\
		\begin{bmatrix}
			\ast & \ast \\
			\ast & \ast 
		\end{bmatrix} &
		\begin{bmatrix}
			\ast &2 \\
			2 & \ast 
		\end{bmatrix} &
		\begin{bmatrix}
		\ast & 3 \\
		3&*
		\end{bmatrix}&
		\begin{bmatrix}
			\ast & \ast \\
			\ast & \ast 
		\end{bmatrix} \\
		\begin{bmatrix}
			\ast & 1 \\
			1 & \ast 
		\end{bmatrix} &
		\begin{bmatrix}
			\ast & \ast  \\
			\ast & \ast 
		\end{bmatrix} &
		\begin{bmatrix}
			\ast & \ast  \\
			\ast & \ast 
		\end{bmatrix}&
		\begin{bmatrix}
		\ast & 4 \\
		4&*
		\end{bmatrix}\\
		\begin{bmatrix}
		\ast & 2 \\
		2 & \ast 
		\end{bmatrix} &
		\begin{bmatrix}
		\ast & \ast  \\
		\ast & \ast 
		\end{bmatrix} &
		\begin{bmatrix}
		\ast & 4  \\
		4 & \ast 
		\end{bmatrix}&
		\begin{bmatrix}
		\ast & * \\
		*&*
		\end{bmatrix}\\
	\end{bmatrix}.
	\label{pda ex 1}
	\end{align}
	The constructed array \( {\bf A}_1 \) is a \((8,8,6,4)\) PDA. \qed
\end{example}
 \begin{thm}
 	\label{thm1}
For a private CDC model with \( K_1 \) nodes and \( K_2 \) output functions, suppose there exist a \((K_1, F_1, Z_1, S_1)\) PDA \({\bf P}^{(1)}\), where each integer appears more than once, and a \((K_2, F_2, Z_2, S_2)\) PDA \({\bf P}^{(2)}\) with \( K_2 > 1 \).
Then, using \textbf{Algorithm 1}, one can construct an extended \((K_1K_2, F_1F_2, Z_1F_2 + (F_1-Z_1)Z_2, S_1S_2)\) PDA \({\bf P}\). This PDA \({\bf P}\) enables a coding scheme for the model that achieves a computation load of 
\begin{align}
	r=\frac{Z_1K_1}{F_1}+ \left (1-\frac{Z_1}{F_1} \right )\frac{Z_2K_1}{F_2}
\end{align} 	and a  communication load of
\begin{align}
	\label{com load}
	L_p= \frac{S_2}{K_1F_1F_2} \left (S_1 +\sum_{g=2}^{K_1} \frac{S_1^g}{ (g-1)}  \right )
\end{align}	where $S_1^g$ is the number of integers in $[S_1]$ which appear exactly $g$ times in the PDA ${\bf P}^{(1)}$.
 \end{thm}
\noindent
Now, we illustrate Theorem \ref{thm1} using a simple example.
\begin{example}
	\label{ex pda 1}
		Suppose we have \(N = 9\) files: \(W_1, W_2, W_3, W_4, W_5, W_6, W_7, W_8, \) and \(W_9\), which are divided into \(F = 9\) batches as follows
	\begin{align}
		B_1 &= \{W_1\}, & B_2 &= \{W_2\}, & B_3 &= \{W_3\}, \nonumber \\  B_4 &= \{W_4\}, & B_5 &= \{W_5\}, & B_6 &= \{W_6\}, \nonumber \\  B_7 &= \{W_7\}, & B_8 &= \{W_8\}, & B_9 &= \{W_9\}.
	\end{align}
	There are \(K = 3\) nodes, and each node computes one output function. Specifically, Node 1 computes \(\phi_1\), Node 2 computes \(\phi_2\), and Node 3 computes \(\phi_3\).
	
	Consider two $2$-$(3,3,1,3)$ PDAs $P^{(1)}$ and $P^{(2)}$ as given by:
	\begin{align}
	P^{(1)} = P^{(2)} =
	\begin{bmatrix}
	\ast & 1 & 2 \\
	1 & \ast & 3 \\
	2 & 3 & \ast
	\end{bmatrix}.
	\end{align}
	An extended PDA \( {\bf P}_1 \) is constructed from \( P^{(1)} \) and \( P^{(2)} \) as follows:
	\begin{itemize}
		\item Replace each integer \(1\) in \(P^{(1)}\) with \(P^{(2)}\), i.e., we have
		$$\begin{bmatrix}
		\ast & 1 & 2 \\
		1 & \ast & 3 \\
		2 & 3 & \ast
		\end{bmatrix}.$$
		\item Replace each integer \(2\) in \(P^{(1)}\) with \(P^{(2)} +3\), i.e., we have 
		$$\begin{bmatrix}
		\ast & 4 & 5 \\
		4 & \ast & 6 \\
		5 & 6 & \ast
		\end{bmatrix}.$$
\item Replace each integer \(3\) in \(P^{(1)}\) with \(P^{(2)} +6\), i.e., we have
$$\begin{bmatrix}
\ast & 7 & 8 \\
7 & \ast & 9 \\
8 & 9 & \ast
\end{bmatrix}.$$
		\item Replace each \(\ast\) in \(P^{(1)}\) with a \(3 \times 3\) block of all \(\ast\)'s:
		\[
		\begin{bmatrix}
		\ast & \ast & \ast \\
		\ast & \ast & \ast \\
		\ast & \ast & \ast
		\end{bmatrix}.
		\]
	\end{itemize}
	The extended $(9,9,5,9)$ PDA constructed using {\bf Algorithm \ref{algo1}} is then represented as:
	\begin{align}
	{\bf P}_1 =
	\begin{bmatrix}
	\begin{bmatrix} \ast & \ast & \ast \\ \ast & \ast & \ast \\ \ast & \ast & \ast \end{bmatrix} &
	\begin{bmatrix} \ast & 1 & 2 \\ 1 & \ast & 3 \\ 2 & 3 & \ast \end{bmatrix} &
	\begin{bmatrix} \ast & 4 & 5 \\ 4 & \ast & 6 \\ 5 & 6 & \ast \end{bmatrix} \\
	\begin{bmatrix} \ast & 1 & 2 \\ 1 & \ast & 3 \\ 2 & 3 & \ast \end{bmatrix} &
	\begin{bmatrix} \ast & \ast & \ast \\ \ast & \ast & \ast \\ \ast & \ast & \ast \end{bmatrix} &
	\begin{bmatrix} \ast & 7 & 8 \\ 7 & \ast & 9 \\ 8 & 9 & \ast \end{bmatrix} \\
	\begin{bmatrix} \ast & 4 & 5 \\ 4 & \ast & 6 \\ 5 & 6 & \ast \end{bmatrix} &
	\begin{bmatrix} \ast & 7 & 8 \\ 7 & \ast & 9 \\ 8 & 9 & \ast \end{bmatrix} &
	\begin{bmatrix} \ast & \ast & \ast \\ \ast & \ast & \ast \\ \ast & \ast & \ast \end{bmatrix}
	\end{bmatrix}.
	\end{align}

\subsubsection*{Data Assignment}
Suppose there are 6 virtual nodes in addition to the 3 real nodes. In total, we have 9 effective nodes (6 virtual  and 3 real nodes).
Each column of the extended PDA corresponds to an effective node, and each row corresponds to a batch of files. Each \(\ast\) in a column indicates that the corresponding batch is stored at that effective node. The extended PDA ${\bf P}_1$ is a block array with 3 row and column blocks. One of the columns in the first column block corresponds to real node 1 and the rest are virtual nodes. Similarly, one of the columns in the second column block corresponds to real node 2, and so on.

	Each real node $k \in [3]$ randomly selects a number $a_k$ from $[3]$ and impersonates the $a_k$-th column from its respective column block. For instance, if $a_1 = 1$, $a_2 = 3$, and $a_3 = 2$, then:  Real node 1 impersonates effective node 1, Real node 2 impersonates effective node 6, and Real node 3 impersonates effective node 8.
	The values of $a_k$ are unknown to the other nodes.
	
	\noindent Each real node stores the following batches:
	\begin{itemize}
		\item {Real node 1:} $B_1, B_2, B_3, B_4, B_7$,
		\item {Real node 2:} $B_3, B_4, B_5, B_6, B_9$,
		\item {Real node 3:} $B_2, B_5, B_7, B_8, B_9$.
	\end{itemize}
	Each node computes the following IVs:
	\begin{itemize}
		\item {Real node 1:} $\{v_{q,n} : q \in [3], n \in \{1,2,3,4,7\}\}$,
		\item {Real node 2:} $\{v_{q,n} : q \in [3], n \in \{3,4,5,6,9\}\}$,
		\item {Real node 3:} $\{v_{q,n} : q \in [3], n \in \{2,5,7,8,9\}\}$.
	\end{itemize}
		The real node $k$ has to compute the function \(\phi_k\), for which the missing IVs are as follows
	\begin{itemize}
		\item {Real node 1:} $\{v_{1,5}, v_{1,6}, v_{1,8}, v_{1,9}\}$,
		\item {Real node 2:} $\{v_{2,1}, v_{2,2}, v_{2,7}, v_{2,8}\}$,
		\item {Real node 3:} $\{v_{3,1}, v_{3,3}, v_{3,4}, v_{3,6}\}$.
	\end{itemize}
	Now, each real node \(k\) generates a random permutation \({\bf y}_k\) of the set \([3]\), with the \(a_k\)-th entry being \(k\), and sends it to all other nodes. The $i$-th entry in the vector \({\bf y}_k\)  represents  the output function index assigned to the effective node $(3(k-1)+i)$.
	For example, if \({\bf y}_1 = \{1, 3, 2\}\), \({\bf y}_2 = \{3, 1, 2\}\), and \({\bf y}_3 = \{2, 3, 1\}\), the demands for each effective node \(j \in [9]\) are represented by the set $D_k$ as follows:
	\begin{align}
	D_1 &= \{v_{1,5},v_{1,6},v_{1,8},v_{1,9}\}, & D_2 &= \{v_{3,4},v_{3,6},v_{3,7},v_{3,9}\}, 
	\nonumber \\
	D_3 &= \{v_{2,4},v_{2,5},v_{2,7},v_{2,8}\}, &
	D_4 &= \{v_{3,2},v_{3,3},v_{3,8},v_{3,9}\}, 	
	\nonumber \\
	D_5 &= \{v_{1,1},v_{1,3},v_{1,7},v_{1,9}\}, & D_6 &= \{v_{2,1},v_{2,2},v_{2,7},v_{2,8}\},
	\nonumber \\
	D_7 &= \{v_{2,2},v_{2,3},v_{2,5},v_{2,6}\}, & D_8 &= \{v_{3,1},v_{3,3},v_{3,4},v_{3,6}\}, 	
	\nonumber \\
	D_9 &= \{v_{1,1},v_{1,2},v_{1,4},v_{1,5}\}.
	\end{align}
	The integers in the PDA represent the demands of the effective nodes. For example, there is a \(1\) in the first column and the fifth row of the PDA ${\bf P}_1$. This indicates that the corresponding effective node demands the IV associated with the files in batch \(B_5\) and the output function \(\phi_1\) (which is the output function assigned to the effective node 1). Therefore, that \(1\) represents the IV \(v_{1,5}\).
	
		Next, we examine the multiplicity of each integer in the first PDA, denoted by $P^{(1)}$. In this array, every integer appears exactly twice. For instance, the integer $1$ occurs in columns 1 and 2 of $P^{(1)}$. Now, corresponding to the integer $1$ in $P^{(1)}$, we associate the integers $1$, $2$, and $3$ in the second PDA, $P^{(2)}$. All demands related to these integers in the extended PDA ${\bf P}_1$ are then renamed with a superscript $2$. This superscript reflects the occurrences of integer $1$ in $P^{(1)}$ and indicates that the associated part of the IV is transmitted as a coded message by real node 2.
			For example, in the extended PDA ${\bf P}_1$, the demand corresponding to integer $1$ in the first column, say $v_{1,5}$, is renamed as $v_{1,5}^2$ to show that this part of the IV is sent by real node 2. In general, the superscript specifies the node responsible for transmitting that particular part of the IV.

		For the integer \(1\) in the first column of \(P^{(1)}\), we consider the integers in PDA \(P^{(2)}\) (i.e., \(1\), \(2\), and \(3\)). Real node \(1\) takes the demands corresponding to the \(1\)s in the columns of column block \(2\) of PDA \(\mathbf{P}_1\), XORs them, and transmits the result. Similarly, real node \(1\) takes the demands corresponding to the \(2\)s and \(3\)s in the columns of column block \(2\) of PDA \(\mathbf{P}_1\) and repeats the same procedure.
		
		Next, we consider the integer \(2\) in the first column of \(P^{(1)}\). For this, we look at the integers in PDA \(P^{(2)} + 3\) (i.e., \(4\), \(5\), and \(6\)). The same procedure is applied to these integers, and so on.
		The transmitted coded symbols are generated as follows
	\begin{align}
	X_1^1 &= v_{3,2}^1 \oplus v_{1,1}^1, \quad X_2^1 = v_{3,3}^1 \oplus v_{2,1}^1, \quad X_3^1 = v_{1,3}^1 \oplus v_{2,2}^1, \nonumber\\
	X_4^1 &= v_{2,2}^1 \oplus v_{3,1}^1, \quad X_5^1 = v_{2,3}^1 \oplus v_{1,1}^1, \quad X_6^1 = v_{3,3}^1 \oplus v_{1,2}^1,\nonumber \\
	X_1^2 &= v_{1,5}^2 \oplus v_{3,4}^2, \quad X_2^2 = v_{1,6}^2 \oplus v_{2,4}^2, \quad X_3^2 = v_{3,6}^2 \oplus v_{2,5}^2, \nonumber\\
	X_7^2 &= v_{2,5}^2 \oplus v_{3,4}^2, \quad X_8^2 = v_{2,6}^2 \oplus v_{1,4}^2, \quad X_9^2 = v_{3,6}^2 \oplus v_{1,5}^2,\nonumber \\
	X_4^3 &= v_{1,8}^3 \oplus v_{3,7}^3, \quad X_5^3 = v_{1,9}^3 \oplus v_{2,7}^3, \quad X_6^3 = v_{3,9}^3 \oplus v_{2,8}^3, \nonumber\\
	X_7^3 &= v_{3,8}^3 \oplus v_{1,7}^3, \quad X_8^3 = v_{3,9}^3 \oplus v_{2,7}^3, \quad X_9^3 = v_{1,9}^3 \oplus v_{2,8}^3.
	\end{align}
	Finally, each node retrieves the necessary IVs to compute its respective output function.
	
	Each real node stores 5 batches of files, resulting in a computation load of \(r = \frac{5 \cdot 3}{9}\). A total of 18 coded symbols are transmitted, each of size \(\alpha\). Therefore, the communication load is \(\frac{18}{3 \cdot 9} = \frac{2}{3}\).
\end{example}
\section{Characterizing Loads for Private CDC Models}
\label{trade}

In this section, we focus on the $g$-regular PDAs constructed using {\textbf{Algorithm \ref{algo2}}}, which correspond to the coded caching problem proposed in \cite{R5}. Using these PDAs, we derive an extended set of PDAs as presented in {\bf Construction \ref{CON1}} below whose proof of correctness  is provided in Appendix \ref{proof cons1}. The achievable computation-communication trade-off points for the private CDC model are then characterized in Theorem \ref{thm2} below, with the proof provided in Appendix \ref{proof thm2}.
{\bf Construction \ref{CON1}} generates a set of extended PDAs. Using these extended PDAs, Theorem \ref{thm2} characterizes the computation-communication trade-off for private CDC. This theorem provides a concise mathematical framework that formalizes {\bf Construction \ref{CON1}}, clarifying its inputs, outputs, and how it enables privacy trade-offs. The PDAs from {\bf Construction \ref{CON1}} enable a private CDC scheme serving K nodes and Q output functions, achieving the specified trade-off in Theorem \ref{thm2}, which summarizes the role of {\bf Construction \ref{CON1}}.
\begin{algorithm}
	\caption{${(t+1)}\text{-}\left (D, {D \choose t},{D-1 \choose t-1}, {D \choose  t+1} \right ) $ PDA construction for some positive integers $D$ and $t$ such that $t \in [D-1] $.}
	\label{algo2}
	\begin{algorithmic}[1]
		\Procedure{{\bf 1}: }{}
		Arrange all subsets of size $ t+1$ from $[D]$ in lexicographic order and for any subset $T'$ of size $t+1$, define $y_{t+1}(T')$ to be its order.
		\EndProcedure \textbf{ 1}
		\Procedure{{\bf 2}: }{}
		Obtain an array  $P_{D,t}$ of size ${D \choose t} \times {D }$.
		Denote the rows by the sets in $\{T\subset [D], |T| = t \}$ and columns by the indices in $\{d: d\in [D]\}$. Define each entry $p_{T,d}$ corresponding to the row $T$ and the column $d$ as
		\begin{align}
		\label{Dk}
		p_{T,d} = \left\{
		\begin{array}{cc}
		*, &  \text{if } |T \cap d| \neq 0 \\
		y_{t+1}(T \cup d), &  \text{if } |T \cap d| = 0 
		\end{array} \right\}.
		\end{align}
		\EndProcedure \textbf{ 2}
	\end{algorithmic}
\end{algorithm}

The following example  demonstrates the construction of $g$-regular PDA using {\bf Algorithm \ref{algo2}}. 
\begin{example}
	Let  $D=5$ and $t=3$. All subsets of size $t +1=4$ from the set $\{1, 2, 3,4,5\}$ are ordered as follows: {$\{ 1,2,3,4\},\{1,2,3,5\},\{ 1,2,4,5\},\{1,3,4,5\}$ and $\{2, 3,4,5\}$}. 
	For each of these subsets, we define the function $y_4(.)$ as follows: $y_4{(1,2,3,4)} = 1, y_4{(1,2,3,5)} = 2, y_4{(1,2,4,5)} = 3,y_4{(1,3,4,5)} = 4, $ and $y_4{(2, 3,4,5)} = 5$. We then construct an array of size ${5 \choose 3} \times 5$ with the columns indexed as $[5]$ and rows indexed as $3$-subset of the set $[5]$. The entries are filled as follows:
	\begin{itemize}
		\item If there is an overlap between row and column indices, we place a $*$ in the corresponding entry.
		\item Otherwise, we take the union of the row index (3-subset) and the column index (single element) forming a 4-subset and place the corresponding value of $y_4(.)$ for that subset as the entry.
	\end{itemize}
	  Following this procedure, we obtain the  array $P_{5,3}$ given below. This array is a $4$-$(5,10,6,5)$ PDA.
\begin{equation}
	\label{eg algo 2 pda}
P_{5,3}=\begin{blockarray}{cccccc}
& \{1\} & \{2\} & \{3\} & \{4\} & \{5\} \\
\begin{block}{c(ccccc)}
\{123\} & * & * & * & 1 & 2  \\
\{124\} & * & * & 1 & * & 3  \\
\{125\} & * & * & 2 & 3 & * \\
\{134\} & * & 1 & * & * & 4 \\
\{135\} & * & 2 & * & 4 & *  \\
\{145\} & * & 3 & 4 & * & * \\
\{234\} & 1 & * & * & * & 5 \\
\{235\} & 2 & * & * & 5 & * \\
\{245\} & 3 & * & 5 & * & * \\
\{345\} & 4 & 5 & * & * & * \\
\end{block}
\end{blockarray}. 
\end{equation}
\end{example}
\noindent We next illustrate the above construction using an example.
\begin{example}
	\label{ex cons 1}
 Define the following arrays 
	\begin{align}
	A^{(3)}= P_{5,3} \text{ (from (\ref{eg algo 2 pda})) and }
	A^{(4)}=
	\begin{bmatrix}
	1 & 2 
	\end{bmatrix}.
	\end{align}
	The array $A^{(3)}$ is a $4$-$(5,10,6,5)$ PDA obtained using {\bf Algorithm \ref{algo2}} with $D=5$ and $t=3$. The array $A^{(4)}$ can be treated as a $(2,1,0,2)$ PDA.
	An extended PDA \( {\bf A}_2 \) is constructed from \( A^{(3)} \) and \( A^{(4)} \)  as follows:
	\begin{itemize}
		\item Replace each integer \(1\) in \(A^{(3)}\) with \(A^{(4)}\), i.e., 	$\begin{bmatrix}
		1 & 2 
		\end{bmatrix}$.
		\item Replace each integer \(2\) in \(A^{(3)}\) with \(A^{(4)} + 2\), i.e., 	$\begin{bmatrix}
		3 & 4 
		\end{bmatrix}$.
		\item Replace each integer \(3\) in \(A^{(3)}\) with \(A^{(4)} + 4\), i.e., 	$\begin{bmatrix}
		5 & 6 
		\end{bmatrix}$.
		\item Replace each integer \(4\) in \(A^{(3)}\) with \(A^{(4)} + 6\), i.e., 	$\begin{bmatrix}
		7 & 8 
		\end{bmatrix}$.
		\item Replace each integer \(5\) in \(A^{(3)}\) with \(A^{(4)} + 8\), i.e., 	$\begin{bmatrix}
		9 & 10 
		\end{bmatrix}$.
		\item Replace each \(\ast\) in $A^{(3)}$ with a \(1 \times 2\) array of all \(\ast\)'s, i.e.,
		$
		\begin{bmatrix}
		\ast & \ast 
		\end{bmatrix}.
		$
	\end{itemize}
	The resulting extended array \({\bf A}_2\) is as given below and it is a  \((10,10,6,10)\) PDA.
	\begin{align}
	{\bf A}_2 =
	\begin{bmatrix}
	\begin{bmatrix}
	\ast & \ast 
	\end{bmatrix} &
	\begin{bmatrix}
	\ast & \ast 
	\end{bmatrix} &
	\begin{bmatrix}
	\ast & \ast 
	\end{bmatrix} &
	\begin{bmatrix}     
	1 & 2 
	\end{bmatrix}&
	\begin{bmatrix}
	3 & 4 
	\end{bmatrix} \\
	\begin{bmatrix}
	\ast & \ast 
	\end{bmatrix} &
	\begin{bmatrix}
	\ast & \ast 
	\end{bmatrix} &
	\begin{bmatrix}
	1 & 2 
	\end{bmatrix} &
	\begin{bmatrix}     
	* & * 
	\end{bmatrix}&
	\begin{bmatrix}
	3 & 4 
	\end{bmatrix} \\
	\begin{bmatrix}
	\ast & \ast 
	\end{bmatrix} &
	\begin{bmatrix}
	\ast & \ast 
	\end{bmatrix} &
	\begin{bmatrix}
	3 & 4 
	\end{bmatrix} &
	\begin{bmatrix}     
	5 & 6 
	\end{bmatrix}&
	\begin{bmatrix}
	* & * 
	\end{bmatrix} \\
	\begin{bmatrix}
	\ast & \ast 
	\end{bmatrix} &
	\begin{bmatrix}
	1 & 2 
	\end{bmatrix} &
	\begin{bmatrix}
	* & * 
	\end{bmatrix} &
	\begin{bmatrix}     
	* & *
	\end{bmatrix}&
	\begin{bmatrix}
	7 & 8 
	\end{bmatrix} \\
	\begin{bmatrix}
	\ast & \ast 
	\end{bmatrix} &
	\begin{bmatrix}
	3 & 4 
	\end{bmatrix} &
	\begin{bmatrix}
	* & * 
	\end{bmatrix} &
	\begin{bmatrix}     
	7 & 8 
	\end{bmatrix}&
	\begin{bmatrix}
	* & * 
	\end{bmatrix} \\
	\begin{bmatrix}
	\ast & \ast 
	\end{bmatrix} &
	\begin{bmatrix}
	5 & 6 
	\end{bmatrix} &
	\begin{bmatrix}
	7 & 8 
	\end{bmatrix} &
	\begin{bmatrix}     
	* & * 
	\end{bmatrix}&
	\begin{bmatrix}
	* & * 
	\end{bmatrix} \\
	\begin{bmatrix}
	1 & 2 
	\end{bmatrix} &
	\begin{bmatrix}
	\ast & \ast 
	\end{bmatrix} &
	\begin{bmatrix}
	* & * 
	\end{bmatrix} &
	\begin{bmatrix}     
	* & * 
	\end{bmatrix}&
	\begin{bmatrix}
	9 & 10
	\end{bmatrix} \\
	\begin{bmatrix}
	3 & 4 
	\end{bmatrix} &
	\begin{bmatrix}
	\ast & \ast 
	\end{bmatrix} &
	\begin{bmatrix}
	* & *
	\end{bmatrix} &
	\begin{bmatrix}     
	9 & 10
	\end{bmatrix}&
	\begin{bmatrix}
	* & * 
	\end{bmatrix} \\
	\begin{bmatrix}
	5 & 6 
	\end{bmatrix} &
	\begin{bmatrix}
	\ast & \ast 
	\end{bmatrix} &
	\begin{bmatrix}
	9 & 10
	\end{bmatrix} &
	\begin{bmatrix}     
	* & *
	\end{bmatrix}&
	\begin{bmatrix}
	* & * 
	\end{bmatrix} \\
	\begin{bmatrix}
	7 & 8
	\end{bmatrix} &
	\begin{bmatrix}
	9 & 10 
	\end{bmatrix} &
	\begin{bmatrix}
	* & *
	\end{bmatrix} &
	\begin{bmatrix}     
	* & *
	\end{bmatrix}&
	\begin{bmatrix}
	* & * 
	\end{bmatrix} \\
	\end{bmatrix}.
	\end{align}
		\qed
\end{example}
 \begin{thm}
	\label{thm2}
		For a private CDC model with \( K \) nodes and \( Q \) output functions, the set of extended PDAs in (\ref{pda 1}), derived from {\bf Construction 1}, corresponds to the model and achieves computation and communication loads given by the points \( (r, L_p) \), where we have
		\begin{align}
			\label{com_load}
		(r, L_p)=	\left(r, \ \frac{Q}{r} \left(1 - \frac{r}{K}\right) \right), \quad r \in [K-1].
	\end{align}
\end{thm}
\begin{remark}
	The computation points achievable for private CDC models, as indicated by Theorem \ref{thm2}, are exactly same as the non-private communication scheme in (\ref{cdc}). The communication points achievable for private CDC models, as indicated by Theorem \ref{thm2}, are scaled by a multiplicative factor of $Q$ compared to the non-private communication scheme in (\ref{cdc}). 
\end{remark}
\noindent Next, we illustrate Theorem \ref{thm2} using an example.
\begin{example}
	\label{ex pda 2}
		We have $3$ nodes and the output functions are assigned as follows: Node 1 computes \(\phi_1\), Node 2 computes \(\phi_2\), and Node 3 computes \(\phi_3\).
	Define the following arrays
	\begin{align}
	P^{(3)}=
	\begin{bmatrix}
	\ast & * & 1 \\
	* & 1 & * \\
	1 & * & \ast
	\end{bmatrix} \text{ and }
	P^{(4)}=
	\begin{bmatrix}
	1 & 2 & 3 
	\end{bmatrix}.
	\end{align}
	The array $P^{(3)}$ is a $3$-$(3,3,2,1)$ PDA while the array $P^{(4)}$ can be treated as a $(3,1,0,3)$ PDA with $K=Q=3$ and $r=2$. The resulting extended PDA \({\bf P}_2\) constructed using {\bf Construction \ref{CON1}} is as given below and it is a  \((9,3,2,3)\) PDA.
	\begin{align}
	{\bf P}_2 =
	\begin{bmatrix}
	\begin{bmatrix}
	\ast & \ast & \ast 
	\end{bmatrix} &
	\begin{bmatrix}
	\ast & \ast & \ast 
	\end{bmatrix} &
	\begin{bmatrix}
	1 & 2 & 3 
	\end{bmatrix} \\
	\begin{bmatrix}
	\ast & \ast & \ast 
	\end{bmatrix} &
	\begin{bmatrix}
	1 & 2 & 3
	\end{bmatrix} &
	\begin{bmatrix}
	\ast & \ast & \ast 
	\end{bmatrix} \\
	\begin{bmatrix}
	1 & 2 & 3
	\end{bmatrix} &
	\begin{bmatrix}
	\ast & \ast & \ast 
	\end{bmatrix} &
	\begin{bmatrix}
	\ast & \ast & \ast 
	\end{bmatrix}
	\end{bmatrix}.
	\end{align} 
		We have \(N = 3\) files: \(W_1, W_2, \) and \(W_3\), divided into \(F = 3\) batches:
	\begin{align}
		B_1 &= \{W_1\}, & B_2 &= \{W_2\}, & B_3 &= \{W_3\}.
	\end{align}

As in Example \ref{ex pda 1}, suppose there are 9 effective nodes: 6 virtual  and 3 real nodes. Each column of the extended PDA corresponds to an effective node, and each row corresponds to a batch of files. A \(\ast\) in a column indicates that the corresponding batch is stored at the respective effective node. The extended PDA \(	{\bf P}_2 \) is a block array with 3 row and column blocks. One column in the first block corresponds to real node 1, and the rest are virtual nodes. Similarly, one column in the second block corresponds to real node 2, and so on.

	Each real node \(k \in [3]\) randomly selects a number \(a_k\) from the set \([3]\) and impersonates the \(a_k\)-th column of block \(k\). For instance, if \(a_1 = 1\), \(a_2 = 2\), and \(a_3 = 3\), then real node 1 impersonates effective node 1, real node 2 impersonates effective node 5, and real node 3 impersonates effective node 9. The values of \(a_k\) are unknown to the other nodes.
	
\noindent	In this set-up, each real node stores batches as follows:
	\begin{align}
		\text{Real node 1:} & \quad \{B_1 ,B_2\}, \nonumber \\
		\text{Real node 2:} & \quad \{B_1 ,B_3 \}, \nonumber \\
		\text{Real node 3:} & \quad \{B_2 ,B_3\}.
	\end{align}
\noindent	Each node computes the following IVs:
	\begin{align}
		\text{Real node 1:} & \quad \{v_{q,n} : q \in [3], n \in \{1,2\}\}, \nonumber\\
		\text{Real node 2:} & \quad \{v_{q,n} : q \in [3], n \in \{1,3\}\}, \nonumber\\
		\text{Real node 3:} & \quad \{v_{q,n} : q \in [3], n \in \{2,3\}\}.
	\end{align}
\noindent	The missing IVs for the real nodes related to their respective output functions are:
	\begin{align}
		\text{Real node 1:} & \quad \{v_{1,3}\}, \nonumber\\
		\text{Real node 2:} & \quad \{v_{2,2}\}, \nonumber\\
		\text{Real node 3:} & \quad \{v_{3,1}\}.
	\end{align}
	Each real node \(k\) generates a random permutation \({\bf y}_k\) of the set \([3]\) with the \(a_k\)-th entry being \(k\), and sends it to all other nodes. The $i$-th entry in the vector \({\bf y}_k\)  represents  the output function index assigned to the effective node $(3(k-1)+i)$. For example, if \({\bf y}_1 = \{ 1,2,3\}\), \({\bf y}_2 = \{3, 2, 1\}\), and \({\bf y}_3 = \{2, 1, 3\}\), the demands for each effective node \(j \in [9]\) are given by the set $D_j$ where we have
	\begin{align}
	D_1 &= \{v_{1,3}\}, & D_2 &= \{v_{2,3}\}, & D_3 &= \{v_{3,3}\},
\nonumber \\
	D_4 &= \{v_{3,2}\}, & D_5 &= \{v_{2,2}\}, & D_6 &= \{v_{1,2}\},
\nonumber \\
	D_7 &= \{v_{2,1}\}, & D_8 &= \{v_{1,1}\}, & D_9 &= \{v_{3,1}\}.
\end{align}
	The integers in the PDA represent the demands of the effective nodes. For example, the integer \(1\) in the first column and the 3rd row of PDA ${\bf P}_2$ indicates that the corresponding effective node demands the IV associated with the files in batch \(B_3\) and the output function \(\phi_1\), represented by \(v_{1,3}\).

Now, we focus on the  PDA $P^{(3)}$, where each integer occurs three times. For the integer $1$ in $P^{(3)}$, we associate it to the integers $1$, $2$, and $3$ in the second PDA, $P^{(4)}$. As a result, all demands corresponding to these integers in the extended PDA ${\bf P}_2$ are divided into two equal parts.
	For instance, since integer $1$ appears in columns 1, 2, and 3 of $P^{(3)}$, the IV $v_{1,3}$ is split into $v_{1,3}^2$ and $v_{1,3}^3$. Here, real node 2 sends $v_{1,3}^2$ as part of its coded multicast, while real node 3 transmits $v_{1,3}^3$. In this way, different nodes are responsible for different portions of the same IV, and the superscripts indicate which node transmits each portion.
	
	The corresponding splitting of the IVs are as follows
	\begin{align}
		v_{1,3} &\!=\! \{v_{1,3}^2,v_{1,3}^3\},  & v_{2,3} &\!=\! \{v_{2,3}^2,v_{2,3}^3\}, & v_{3,3} &\!=\! \{v_{3,3}^2,v_{3,3}^3\},  \nonumber\\
		v_{3,2} &\!=\! \{v_{3,2}^1,v_{3,2}^3\}, 
		&	v_{2,2} &\!=\! \{v_{2,2}^1,v_{2,2}^3\}, 
		&v_{1,2} &\!=\! \{v_{1,2}^1,v_{1,2}^3\}, \nonumber\\
		v_{2,1} &\!=\! \{v_{2,1}^1,v_{2,1}^2\}, 
		&
		v_{1,1} &\!=\! \{v_{1,1}^1,v_{1,1}^2\}, 
		&
		v_{3,1} &\!=\! \{v_{3,1}^1,v_{3,1}^3\}.
	\end{align}
	For integer 1 present in the first column of PDA \(P^{(3)}\), we consider the integers in PDA \(P^{(4)}\) (i.e., 1, 2, and 3).  Real node 1 takes the parts of the demands corresponding to the 1s in the columns of column blocks 2 and 3 of PDA ${\bf P}_2$ with a superscript of 1, XORs them, and sends them. Similarly real node 1 takes the parts of the demands corresponding to the 2s and 3s in the columns of column blocks 2 and 3 of PDA ${\bf P}_2$ and repeat the same procedure.
	Real nodes 2 and 3 follow the same procedure.
	The transmitted  coded symbols are
	\begin{align}
	X_1^1 &= v_{3,2}^1 \oplus v_{2,1}^1,&
	X_2^1 &= v_{2,2}^1 \oplus v_{1,1}^1,&
	X_3^1 &= v_{3,1}^1 \oplus v_{1,2}^1, \nonumber \\
X_1^2 &=v_{1,3}^2 \oplus v_{2,1}^2,&
X_2^2 &= v_{2,3}^2 \oplus  v_{1,1}^2,&
X_3^2 &= v_{3,3}^2 \oplus  v_{3,1}^2,\nonumber\\
X_1^3 &= v_{1,3}^3,\oplus v_{3,2}^3, &
X_2^3 &= v_{2,3}^3 \oplus v_{2,2}^3,&
X_3^3 &= v_{3,3}^3 \oplus v_{1,2}^3.
\end{align}
	Finally, all nodes retrieve whatever is required to compute their respective output functions.
	
	Each real node stores 2 batches of files. The computation load is \(r = 2\). A total of 9 symbols are transmitted, each of size \(\frac{\alpha}{2}\). Hence, the communication load is \(\frac{9}{ 2 \cdot 3 \cdot 3} = \frac{1}{2}\).
\end{example}

\section{Improved Computation-Communication Trade-offs}
\label{opt}
In Section \ref{trade}, we observed that while the computation loads achieved in the private scenario remain the same as those in the non-private case, the communication load increases by a factor of \( Q \). In this section, we construct a different set of extended PDAs that allow one to achieve a reduced communication load. 

As before, we focus on \( g \)-regular PDAs generated using \textbf{Algorithm \ref{algo2}}. By utilizing these PDAs, we derive another extended set of PDAs, as outlined in \textbf{Construction \ref{CON2}} below. The proof of correctness of \textbf{Construction \ref{CON2}} is provided in Section \ref{proof cons2}. The improved computation-communication trade-off points achievable for the private CDC model are then given in {Theorem \ref{thm3}}, with the proof provided in {Section \ref{proof thm3}}.

\begin{constr}
	\label{CON2}
	Consider the following two sets of PDAs
	\begin{align}
	\mathcal{P}^{(1)} \!=\!\Biggl \{(r_1+1)\text{-}\left ({K}, {K \choose r_1},{K-1 \choose r_1-1},  {K \choose  r_1+1} \right )  \text{ PDA}: \nonumber \\r_1 \in [K-1] \Biggr \}
	\end{align} 
	\begin{align}
	\mathcal{P}^{(2)} \!=\!	\Biggl \{(r_2+1)\text{-}\left ({Q}, {Q \choose r_2},{Q-1 \choose r_2-1},  {Q \choose  r_2+1} \right )  \text{ PDA}: 
	\nonumber \\r_2 \in [Q-1] \Biggr \}
	\end{align} 
	constructed using {\textbf{Algorithm \ref{algo2}}}, for some positive integers $K, Q, r_1$ and $r_2$ such that $Q \geq K$. From this set of PDAs we construct another set of extended PDAs  using {\textbf{Algorithm \ref{algo1}}} as follows
	\begin{align}
	\label{pda 2}
	\mathcal{P} =	\Biggl \{\left (K,F,Z,S \right )  \text{ PDA}: K={KQ}, F={K \choose r_1}{Q \choose r_2}, \nonumber\\Z={K-1 \choose r_1-1}{Q \choose  r_2+1} \left(\frac{Q}{r_2} -\frac{K-r_1}{r_1} \right), \nonumber\\S= {K \choose  r_1+1} {Q \choose  r_2+1} ,r_1 \in [K-1],r_2 \in [Q-1] \Biggr \}.
	\end{align} 
\end{constr}
\begin{thm}
	\label{thm3}
	For a private CDC model with \( K \) nodes and \( Q \) output functions, the set of PDAs in (\ref{pda 2}), obtained from {\bf Construction 2}, corresponds to the model and achieves computation and communication loads given by the points \( (r, L_p) \), where we have
	\begin{align}
(r, L_p)=	\left (r_1+ \frac{(K-r_1)r_2}{Q},\frac{1}{r_1} \left ( 1-\frac{r_1}{K}\right )\left ( \frac{Q-r_2}{r_2+1} \right) \right ), \nonumber \\ \forall r_1 \in [K-1], r_2 \in [Q-1].
	\label{L for}
\end{align} 
\end{thm}
One extreme scenario occurs when  \( r_1 = r_2 = 1 \) as in Example \ref{ex pda 1}. In this case, if we treat the PDAs constructed using \textbf{Algorithm \ref{algo1}} as block arrays, all the elements in the diagonal blocks are stars, and for all other blocks only one star is present in each of the columns. Here, the computation load is the lowest among the PDAs constructed using {\bf Construction \ref{CON2}} (since the least number of stars are present in each column). The computation load in this scenario is \( r = 1 + \frac{K-1}{Q} \). The communication load is \( L_p = \left(1 - \frac{1}{K}\right) \frac{Q-1}{2} \).

The other extreme happens when  \( r_1 = K-1 \) and \( r_2 = Q-1 \), in which case the PDA is an array of size \( KQ \times KQ \), with the anti-diagonal entries being 1 and all other entries being \( * \). In this case, the computation load is the highest among the PDAs constructed using {\bf Construction \ref{CON2}} (since only one integer appears in the PDA). The computation load for this scenario is \( r = K - \frac{1}{Q} \) . The communication load is given by \( L_p = \frac{1}{(K-1)KQ} \).

Note that for PDAs constructed using {\bf Construction \ref{CON2}}, the minimum computation load is \( r = 1 + \frac{K-1}{Q} \), which never reaches 1. Conversely, the maximum computation load is \( r = K - \frac{1}{Q} \), which exceeds \( K - 1 \).
\begin{remark}
	The computation points achievable for private CDC models, as described in \textbf{Theorem \ref{thm3}}, are shifted by an additive factor of \( \frac{(K-r_1)r_2}{Q} \) compared to the non-private communication scheme in \((\ref{cdc})\). Similarly, the communication points achievable for private CDC models, as stated in \textbf{Theorem \ref{thm2}}, are scaled by a multiplicative factor of \( \left(\frac{Q-r_2}{r_2+1}\right) \) relative to the non-private communication scheme in \((\ref{cdc})\). \qed
\end{remark}
From Theorem \ref{thm2}, the achievable computation and communication loads are represented by a piecewise linear curve with corner points given by:
$
\left( r, \frac{Q}{r} \left(1 - \frac{r}{K}\right) \right), \quad r \in [K-1].
$
Thus, the communication load achievable at any point \( r \) is:
$
L_p(r) = \frac{Q}{r} \left(1 - \frac{r}{K}\right).
$
From Theorem \ref{thm3}, the achievable computation and communication loads are represented by another piecewise linear curve with corner points:
$
\left( r_1 + \frac{(K - r_1) r_2}{Q}, \frac{1}{r_1} \left(1 - \frac{r_1}{K}\right) \left(\frac{Q - r_2}{r_2 + 1}\right) \right),  r_1 \in [K-1], r_2 \in [Q-1].
$
The communication load achievable using Theorem \ref{thm2} at the point \( r_1 + \frac{(K - r_1) r_2}{Q} \) is
\begin{align}
A &\overset{\Delta}{=} L_p(r_1+ \frac{(K-r_1)r_2}{Q}) \nonumber \\&= \frac{Q}{r_1 + \frac{(K - r_1) r_2}{Q}} \left(1 - \frac{r_1 + \frac{(K - r_1) r_2}{Q}}{K}\right)
\nonumber \\&= \frac{Q^2}{r_1 Q + (K - r_1) r_2}\left(1 - \frac{r_1 Q + (K - r_1) r_2}{K Q}\right)
\nonumber \\&= \frac{Q^2}{r_1 Q + (K - r_1) r_2}\left(\frac{(K - r_1)(Q - r_2)}{K Q}\right)
\nonumber \\ &= \frac{Q (K - r_1)(Q - r_2)}{K [r_1 Q + (K - r_1) r_2]}.
\end{align}
On the other hand, the communication load achievable using Theorem \ref{thm3} at the same point is:
\begin{align}
B &\overset{\Delta}{=} \frac{1}{r_1} \left(1 - \frac{r_1}{K}\right) \left(\frac{Q - r_2}{r_2 + 1}\right)
\nonumber \\ &= \frac{(K - r_1)(Q - r_2)}{K r_1 (r_2 + 1)}.
\end{align}
To compare \( A \) and \( B \), divide \( A \) by \( B \):
\begin{align}
\frac{A}{B} &	= \frac{Q (K - r_1)(Q - r_2)}{K [r_1 Q + (K - r_1) r_2]} \times \frac{K r_1 (r_2 + 1)}{(K - r_1)(Q - r_2)}
\nonumber \\ &= \frac{Q r_1 (r_2 + 1)}{r_1 Q + (K - r_1) r_2}.
\end{align}
We now prove by contradiction that \( A > B \). Assume \( A \leq B \), which implies
\begin{align}
\frac{Q (r_2 + 1)}{Q - r_2 + \frac{K r_2}{r_1}} &\leq 1
\nonumber \\ \Rightarrow  Q (r_2 + 1) &\leq Q - r_2 + \frac{K r_2}{r_1}
\nonumber \\ \Rightarrow  Q r_2 &\leq - r_2 + \frac{K r_2}{r_1}
\nonumber\\ \Rightarrow  r_2 (Q + 1) &\leq \frac{K r_2}{r_1}
\nonumber \\\Rightarrow  r_1 (Q + 1) &\leq K.
\end{align}
This contradicts our assumption that \( Q \geq K \), since \( r_1 \geq 1 \) and \( Q + 1 > K \). Therefore, the assumption \( A \leq B \) is false, and we conclude that \( A > B \).

We saw one extreme scenario when \( r_1 = r_2 = 1 \) in Example \ref{ex pda 1}. We next illustrate the other extreme scenario when \( r_1 = K-1 \) and \( r_2 = Q-1 \)  through an example.

\begin{figure}[t]
	\centering
	\includegraphics[scale=0.34]{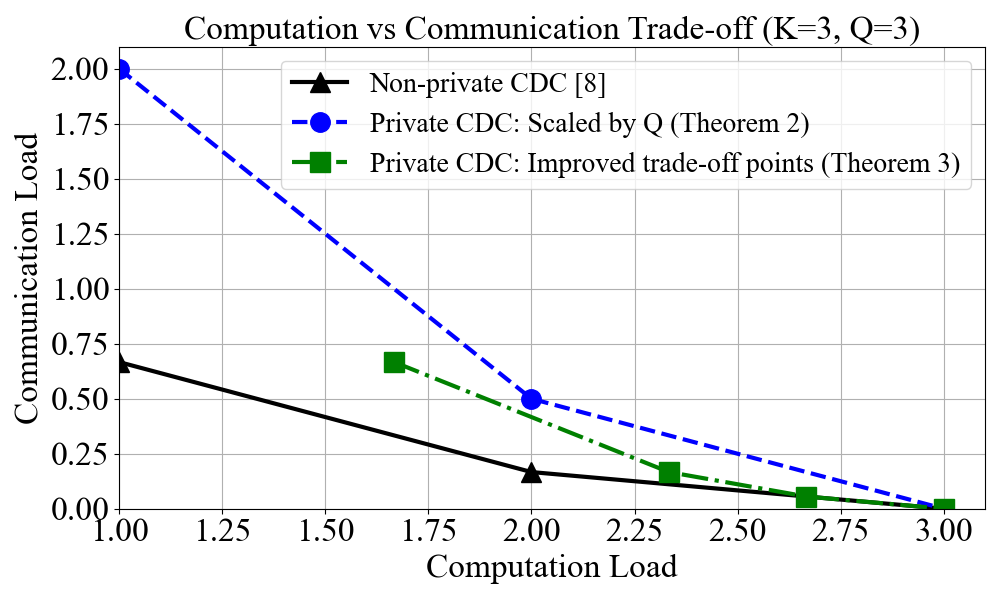}
	\caption{Comparison of computation–communication trade-offs for non-private and private CDC schemes. Here, the number of nodes considered is \( K = 3 \) and output functions is \( Q=3 \).}
	\label{fig:tradeoff}
\end{figure}

	\begin{example}
		\label{ex algo 3}
			We consider a  setup consisting of $N = 9$ files, grouped into $F = 9$ batches $\{B_f: f \in [9]\}$, with output functions assigned as follows: Node 1 computes $\phi_1$, Node 2 computes $\phi_2$, and  Node 3 computes $\phi_3$.
		The two given PDAs in  {\bf Algorithm \ref{algo1}} are defined as
		\begin{align}
		P^{(5)} = P^{(6)} =
		\begin{bmatrix}
		\ast & \ast & 1 \\
		\ast & 1 & \ast \\
		1 & \ast & \ast
		\end{bmatrix}.
		\end{align}
		Both $P^{(5)}$ and $P^{(6)}$ are $3$-$(3,3,2,1)$ PDAs.
		The resulting extended array ${\bf P}_3$ is
		\begin{align}
		{\bf P}_3 =
		\begin{bmatrix}
		\begin{bmatrix}
		\ast & \ast & \ast \\
		\ast & \ast & \ast \\
		\ast & \ast & \ast
		\end{bmatrix} &
		\begin{bmatrix}
		\ast & \ast & \ast \\
		\ast & \ast & \ast \\
		\ast & \ast & \ast
		\end{bmatrix} &
		\begin{bmatrix}
		\ast & \ast & 1 \\
		\ast & 1 & \ast \\
		1 & \ast & \ast
		\end{bmatrix} \\
		\begin{bmatrix}
		\ast & \ast & \ast \\
		\ast & \ast & \ast \\
		\ast & \ast & \ast
		\end{bmatrix} &
		\begin{bmatrix}
		\ast & \ast & 1 \\
		\ast & 1 & \ast \\
		1 & \ast & \ast
		\end{bmatrix} &
		\begin{bmatrix}
		\ast & \ast & \ast \\
		\ast & \ast & \ast \\
		\ast & \ast & \ast
		\end{bmatrix} \\
		\begin{bmatrix}
		\ast & \ast & 1 \\
		\ast & 1 & \ast \\
		1 & \ast & \ast
		\end{bmatrix} &
		\begin{bmatrix}
		\ast & \ast & \ast \\
		\ast & \ast & \ast \\
		\ast & \ast & \ast
		\end{bmatrix} &
		\begin{bmatrix}
		\ast & \ast & \ast \\
		\ast & \ast & \ast \\
		\ast & \ast & \ast
		\end{bmatrix}
		\end{bmatrix}.
		\label{pda ex 1}
		\end{align}
		The constructed PDA \( {\bf P}_3 \) is a \((9,9,8,1)\) PDA.

We assume that we have 9 effective nodes (6 virtual  and 3 real nodes). Each column in the extended PDA corresponds to an effective node, while each row represents a batch of files. An asterisk ($\ast$) in a column indicates that the corresponding batch is stored at that effective node. The extended PDA $\mathbf{P}_3$ is divided into 3 row and column blocks, with one column from each block corresponding to a real node.
		
		Each real node \(k \in [3]\) randomly selects a number \(a_k\) from the set \([3]\) and impersonates the \(a_k\)-th column from the column block \(k\). For example, if \(a_1 = 2\), \(a_2 = 3\), and \(a_3 = 2\), then real node 1 impersonates effective node 2, real node 2 impersonates effective node 6, and real node 3 impersonates effective node 8. The value of \(a_k\) is unknown to all other nodes. 
		
		\noindent Real nodes store batches as follows:
		\begin{align}
		\text{Real node 1:} &\quad \{B_f: f \in [9]\backslash 8\}, \nonumber \\ \text{Real node 2:}& \quad \{B_f: f \in [9]\backslash 4\}, \nonumber \\ 	\text{Real node 3:} &\quad \{B_f: f \in [9]\backslash 2\}
		\end{align} 
		and each real node computes the following IVs:
		\begin{align}
		\text{Real node 1: } & \{v_{q,n} : q \in [3], n \in [9]\backslash 8\}, \nonumber \\ \text{ Real node 2: } & \{v_{q,n} : q \in [3], n \in [9]\backslash 4\}, \nonumber \\ \text{Real node 3: } & \{v_{q,n} : q \in [3], n \in [9]\backslash 2\}.
		\end{align}
		The real node 1 has to compute the function \(\phi_1\), for which the missing IV is \(\{v_{1,8}\}\). Similarly, the missing IV for real node 2 is \(\{v_{2,4}\}\), and for real node 3 is \(\{v_{3,2}\}\).
		
		Now, each real node \(k\) generates a random permutation \({\bf y}_k\) of the set \([3]\) with the \(a_k\)-th entry being \(k\), and sends it to all other nodes. For example, if \({\bf y}_1 = \{2, 1, 3\}\), \({\bf y}_2 = \{1, 3, 2\}\), and \({\bf y}_3 = \{1, 3, 2\}\), then the demands of the effective node $j$ appear as $D_j$, where we have
		\begin{align}
		D_1 = \{v_{2,9}\},  \quad D_2 = \{v_{1,8}\},  \quad D_3 = \{v_{3,7}\},
		\nonumber \\
		D_4 = \{v_{1,6}\},  \quad D_5 = \{v_{3,5}\},  \quad D_6 = \{v_{2,4}\},
		\nonumber \\
		D_7 = \{v_{1,3}\},  \quad D_8 = \{v_{3,2}\},  \quad D_9 = \{v_{2,1}\}.
		\end{align}
		The integers in the PDA represent the demands of the effective nodes. For example, there is a \(1\) in the first column and the ninth row of the PDA ${\bf P}_3$. This indicates that the corresponding effective node demands the IV associated with the files in batch \(B_9\) and the output function \(\phi_2\) (which is the output function assigned to the effective node 1). Therefore, that \(1\) represents the IV \(v_{2,9}\).
		
		Since there is only one integer in the PDA $P^{(5)}$ and it occurs three times, we split all the IVs into two equal parts. The superscripts are assigned in such a way that, for example, an integer \(1\) appears in columns 1, 2, and 3 of the PDA $P^{(5)}$. The demand corresponding to the integer \(1\) in the first column, i.e., \(v_{2,9}\), is split into two equal parts with superscripts \(2\) and \(3\), since the other two columns in which \(1\) appears in $P^{(5)}$ are columns 2 and 3. We have
		\begin{align}
		v_{2,9} &= \{v_{2,9}^2, v_{2,9}^3\}, & v_{1,8} &= \{v_{1,8}^2, v_{1,8}^3\}, & v_{3,7} &= \{v_{3,7}^2, v_{3,7}^3\}, \nonumber
		\\
		v_{1,6} &= \{v_{1,6}^1, v_{1,6}^3\}, & v_{3,5} &= \{v_{3,5}^1, v_{3,5}^3\}, & v_{2,4} &= \{v_{2,4}^1, v_{2,4}^3\}, \nonumber
		\\
		v_{1,3} &= \{v_{1,3}^1, v_{1,3}^2\}, & v_{3,2} &= \{v_{3,2}^1, v_{3,2}^2\}, & v_{2,1} &= \{v_{2,1}^1, v_{2,1}^2\}.
		\end{align}
		Now, the real node 1 takes the parts of the demands corresponding to the 1s in the columns of column blocks 2 and 3 with a superscript \(1\), XORs them, and sends them to other real nodes. Similarly, real node 2 takes the parts of the demands corresponding to the 1s in the columns of column blocks 1 and 3 with a superscript \(2\), XORs them, and sends them to other real nodes, and so on.
		The transmitted coded symbols are
		\begin{align}
		X_1^1 &= v_{1,6}^1 \oplus v_{3,5}^1 \oplus v_{2,4}^1 \oplus v_{1,3}^1 \oplus v_{3,2}^1 \oplus v_{2,1}^1,
		\nonumber \\
		X_1^2 &= v_{2,9}^2 \oplus v_{1,8}^2 \oplus v_{3,7}^2 \oplus v_{1,3}^2 \oplus v_{3,2}^2 \oplus v_{2,1}^2,
		\nonumber \\
		X_1^3 &= v_{2,9}^3 \oplus v_{1,8}^3 \oplus v_{3,7}^3 \oplus v_{1,6}^3 \oplus v_{3,5}^3 \oplus v_{2,4}^3.
		\end{align}	
		Real node 1 retrieves \(v_{1,8}^2\) from \(X_1^2\) and \(v_{1,8}^3\) from \(X_1^3\). Similarly, all other nodes retrieve the required data to compute their respective output functions.
		
		Each real node stores 8 batches of files. The computation load is \(r = \frac{8 \cdot 3}{9} = 2.66\). A total of 3 coded symbols are transmitted, each of size \(\frac{\alpha}{2}\). Hence, the communication load is \(\frac{3}{2 \cdot 3 \cdot 9} = \frac{1}{18}\).
		
	\end{example}
To illustrate the trade-offs in Theorems 2 and 3, Fig.~\ref{fig:tradeoff} compares the achievable computation–communication points of the proposed private schemes with the non-private optimal baseline. For a system with $K=3$ nodes and $Q=3$ output functions, the private schemes in Theorem 2 results in an increase in communication by a factor of 3. Moreover, the trade-off is observed to be improved using schemes in Theorem 3.
\section{Discussion}
The proposed extended PDA framework provides theoretical guarantees for private CDC, yet its practical implementation faces challenges in computational overhead and scalability. The total number of files required grows multiplicatively as $F = F_1F_2$, where $F_1 = \binom{K}{r}$ and $F_2$ depends on $Q$, leading to combinatorial scaling with $K$ and $Q$. For instance, in \textbf{Construction 1}, $F = \binom{K}{r}$, while in \textbf{Construction 2}, $F = \binom{K}{r_1}\binom{Q}{r_2}$. Encoding involves XOR operations over IVs mapped to PDA entries. The complexity scales polynomially with $S$, where $S = Q\binom{K}{r+1}$ in  \textbf{Construction 1} and  $S = \binom{K}{r_1+1}\binom{Q}{r_2+1}$ in \textbf{Construction 2} corresponds to the number of coded symbols. Decoding requires solving linear systems proportional to the communication load, further increasing computational demands. Scalability is compounded by the use of $K(Q-1)$ virtual nodes, expanding the effective system size to $KQ$, and the need for $\mathcal{O}(Q!)$ permutations to preserve privacy. Mitigation strategies include hierarchical designs to partition large networks, structured PDAs to limit the number of files required, and approximate schemes trading partial privacy for complexity reduction. 
Addressing these challenges requires future work on hybrid approach, and sparsity-aware PDA designs. Exploring and formalizing these mitigation strategies remain an important direction for future work.

\appendices
  \section{Proof of Correctness of {\bf Algorithm \ref{algo1}}}
\label{proof algo}
The array ${\bf P}$ is a block array with $F_1$ row blocks indexed by $[F_1]$ and ${K}_1$ column blocks indexed by $[{K}_1]$. The block entries are represented by $\{{\bf p}_{f,k}: f \in [F_1] , k \in [K_1]\}$.
The entries include the blocks $ {\bf X}$ and ${\bf P}_s={\bf P}^{(2)} +(s-1)S_2$, where $* + (s-1)S_2 = *$ and $s\in [S_1]$, each of size $F_2 \times K_2$. Hence, the overall size of the array ${\bf P}$ is $F_1F_2 \times K_1K_2$. There are $Z_1$ stars in each  column of the PDA ${\bf P}^{(1)}$ and each star is replaced by ${\bf X}$ where there are $F_2$ stars in each column of ${\bf X}$. Moreover, there are $(F_1-Z_1)$ integers in each column of the PDA ${\bf P}^{(1)}$ and each integer $s \in [S_1]$ is replaced by the array  ${\bf P}_s$ where there are $Z_2$ stars in each column of  ${\bf P}_s$. Consequently, each column of the array ${\bf P}$ contains $Z_1F_2 + (F_1-Z_1)Z_2$ stars and hence, condition $A1$ in Definition \ref{def:PDA} is satisfied. The set of all integers present in the block ${\bf P}_s$ lies in the range $[(s-1)S_2+1,sS_2]$ for each $s \in S_1$. Overall, there are a total of $S_1S_2$ integers in the range $[S_1S_2]$ within the array ${\bf P}$ and, therefore, condition $A2$ in Definition \ref{def:PDA} is satisfied. Notice that for two distinct block entries ${\bf p}_{f_1,k_1}$ and ${\bf p}_{f_2,k_2}$ from  {\textbf{Algorithm \ref{algo1}}}, ${\bf p}_{f_1,k_1}={\bf p}_{f_2,k_2} = {\bf P}_s$ for some $s \in [S_1]$ only if $f_1 \neq f_2$ and $k_1 \neq k_2$. They lie in distinct row and column blocks. Additionally, ${\bf p}_{f_1,k_2} = {\bf p}_{f_2,k_1} = {\bf X},$ i.e., the corresponding $2 \times 2$ sub-block array formed by row blocks $f_1, f_2$ and column blocks $k_1, k_2$ must be either of the following forms
$ \begin{pmatrix}
	{\bf P}_s& {\bf X}\\
	{\bf X} & {\bf P}_s
\end{pmatrix} $or 
$\begin{pmatrix}
	{\bf X}& {\bf P}_s\\
	{\bf P}_s & {\bf X}
\end{pmatrix}.$   Note that for any two distinct values $s_1$ and $s_2$, where $s_1,s_2 \in [S_1]$, the integers present in the blocks ${\bf P}_{s_1}$ and ${\bf P}_{s_2}$ are distinct. Hence, each entry in ${\bf P}_s$, for $s \in [S_1]$, satisfies condition A3 in Definition \ref{def:PDA}. Therefore, we conclude that the array ${\bf P}$ is a $(K_1K_2,F_1F_2,Z_1F_2 + (F_1-Z_1)Z_2,S_1S_2)$  PDA.
\section{Proof of Theorem \ref{thm1}}
\label{proof thm1}
Based on a $(K_1K_2,F_1F_2,Z_1F_2 + (F_1-Z_1)Z_2,S_1S_2)$ PDA ${\bf P}$ obtained using {\textbf{Algorithm \ref{algo1}}}, a private CDC scheme for a model having  $K_1$  nodes can be obtained as given below where we consider $Q=K_2$ output functions.
\subsection{Map Phase}
First, files are divided by grouping $N$ files into $F=F_1F_2$ disjoint batches $\{B_{1},B_{2},\ldots, B_{F_1F_2}\}$, each containing $ \eta =  \frac{N}{F_1F_2}$ files such that $\bigcup_{m \in [F_1F_2]} B_{m} = \{W_1,W_2,\ldots,W_{N}\}$. As described in Appendix \ref{proof algo}, the PDA ${\bf P}$ is an array  with $F_1F_2$ rows indexed by $[F_1F_2]$ and $K_1K_2$ column indexed by $[K_1K_2]$ with the entries represented by $\{{ p}_{f,j}: f \in [{ F_1F_2}] , j \in [{ K_1K_2}]\}$. The rows in the PDA represent the $F$ batches $\{B_{1},B_{2},\ldots, B_{F}\}$.

We assume that there are $K_1K_2$ effective nodes involved in the system denoted by $[K_1K_2]$, out of which $K_1$ nodes are the real nodes and the other $ (K_2-1)K_1$ nodes are the virtual nodes. The columns of the PDA represent the $K_1K_2$ effective nodes. Each effective node $j \in [K_1K_1]$ is assigned a subset of batches as follows:
$ \M_j = \{B_{f} : p_{f,j} = *, f \in [F]\}.$ 

The array ${\bf P}$ is  a block array with $F_1$ row blocks and ${K}_1$ column blocks with entries being the blocks $ {\bf X}$ and ${\bf P}_s={\bf P}^{(2)} +(s-1)S_2$, where $* + (s-1)S_2 = *$ and $s\in [S_1]$, each of size $F_2 \times K_2$.  For each ${ k} \in [{ K_1}]$, the $K_2$ columns in the column block ${ k} $ in the PDA ${\bf P}$ represent the effective nodes in the range $[({ k} -1)K_2+1,{ k} K_2]$. One of the columns in the column block $k$ represents the real node $k$ and the rest of the $(K_2-1)$ columns represent the virtual nodes.

Each real node, denoted by  $k\in [K_1]$, independently selects a value $a_k$ from the set $[K_2]$ uniformly at random. The real node $k$  impersonates  the effective node $\beta_k := (k-1)K_2+ a_k$, representing the column indexed by $\beta_k$ in the PDA ${\bf P}$ ($a_k^{th}$ column in the column block $k$). Importantly, the realization of $a_k$ remains unknown to all the other real nodes, meaning they do not have information about the specific value of $\beta_k \in [(k-1)K_2+1,kK_2]$. Each real node $k \in [K_1]$ stores all the files in the set $\M_{\beta_k} = \{B_{f} : p_{f,\beta_k} = *, f \in [F]\}$.  Hence, it can retrieve  the following IVs
\begin{align}
	\label{IVs}
	V_{\beta_k}= \{v_{q,n} : q \in [K_2], W_n \in B_{f},  B_{f} \in \M_{\beta_k}, f\in [F]\}
\end{align}
where $V_{\beta_k}$ is the set of IVs that can be computed from the files accessible to the real node $k$.

\subsection{Shuffle Phase}
For each real node $k\in [K_1]$ assigned to compute the output function indexed by  $d_k\in[K_2]$ (i.e., the output function $\phi_{d_k}$), the following steps occur

\begin{itemize}
	\item Real node $k$ selects a vector $ {\bf y}_k =(y_{(k-1)K_2+1}, \ldots, y_{ k K_2 } ) $ from all permutations of  $[K_2]$ such that the $a_k$-th element is equal to $d_k$. The $i$-th entry in the vector \({\bf y}_k\)  represents  the output function index assigned to the effective node $((k-1)K_2+i)$.
	\item Real node $k$ broadcasts ${\bf y}_k$ to all the other real nodes. As a result, from the perspective of all other real nodes, the union of the demands from the effective nodes in  the set $ [(k-1)K_2+1,kK_2]$ always covers the entire set of output function indices $[K_2]$, which is crucial for ensuring privacy.
\end{itemize}  

Next, consider the PDA ${\bf P}^{(1)}$ (with the entries represented by $p^1_{i,k}, i \in [F_1], k \in [K_1]$). For each pair $(i, k) \in [F_1] \times [K_1]$ such that
$p_{i,k}^1 = s \in [S_1]$, let $g_s$ be the number of occurrences of $s$. Assume that the other $g_s-1$ occurrences of $s$ are $p^1_{i_1,k_1} = p^1_{i_2,k_2} = \ldots = p^1_{i_{g_s-1},k_{g_s-1}} = s. $
Now, from the construction of the PDA ${\bf P}$, consider the integers in the set  $[(s-1)S_2+1, $ $sS_2]$ from the block ${\bf P}_s$ and the set of all effective nodes in the set $\{[(k_l-1)K_2+1,k_lK_2]:  k_l \in \{k_1,k_2,\ldots,k_{g_s-1}\}\}$. Note that the integers in the set $[(s-1)S_2+1, $ $ sS_2]$ are only present in the columns indexed by $\{[(k_l-1)K_2+1,k_lK_2]:  k_l \in k \cup \{k_1,k_2,\ldots,k_{g_s-1}\}\}$ (from the construction of ${\bf P}$).  For each $t \in [(s-1)S_2+1, $ $sS_2]$, let us consider the set of ordered pairs $\J_t^k =   \{(f,j): p_{f,j}=t, {f} \in [F],j \in \{[(k_l-1)K_2+1,k_lK_2]:  k_l \in \{k_1,k_2,\ldots,k_{g_s-1}\}\}\}$.

Let $T_{k}$ represent the set of all integers present in the $\beta_k^{th}$ column of ${\bf P}$. Let $p_{f,\beta_k} = t \in T_{k}$. We concatenate the set of IVs for the output function $\phi_{y_{{\beta_k}}}(=\phi_{d_k})$ that needs to be computed by the effective node $\beta_k$ (real node $k$) and can be computed from the files in $B_{{f}}$, i.e., $\{v_{d_k,n} :  W_n \in B_{{f}} \}$, into the symbol 
\begin{align}
	\label{symbols_pda}
	\U_{d_k,B_{{f}}} = (v_{d_k,n} :  W_n \in B_{{f}}) \in \F_{2^{\eta \alpha}}.
\end{align}
We partition the symbols in $\U_{d_k,B_{{f}}}$ into $g_s-1$ packets, each of equal size, i.e., we have
\begin{align}
	\label{partition symbols}
	\U_{d_k,B_{{f}}}=\{\U_{d_k,B_{{f}}}^{k_1}, \U_{d_k,B_{{f}}}^{k_2}, \ldots, \U_{d_k,B_{{f}}}^{k_{g_s-1}}\}.
\end{align}	Similarly, for each $(f,j) \in \J_t^k$, we concatenate the set of IVs for the output function $\phi_{y_{j}} $ which are assumed to be computed by the effective node $j$ and can be computed from the files in $B_{{f}}$, into the symbol $\U_{y_j,B_{{f}}}$ and 
we partition the symbols in $\U_{y_j,B_{{f}}}$ into $g_s-1$ packets of equal sizes.

For each $k \in [K_1]$, let $C_{k}$ be the set of integers present in the column indexed by $k$ in PDA ${\bf P}^{(1)}$. For each entry $s \in C_{k}$ and $t \in [(s-1)S_2+1,sS_2]$, the real node $k$ creates a coded symbol
\begin{align}
	\label{transmission}
	X_t^k = \bigoplus_{({f},{j}) \in \J_t^k} \U_{{y_{{j}}},B_{{f}}}^{k}
\end{align} and multicasts the sequence 
$	{\bf X}_k = \{X_t^k: t \in [(s-1)S_2+1,$ $ sS_2],s \in C_k\}.$
The real node $k$ can create the coded symbol $X_t^k$ from the IVs accessible to it. This follows since, for each column block indexed by $k_l \in \{k_1,k_2,\ldots,k_{g_s-1}\}$, the $2 \times 2$ sub-block array formed by row blocks $i, i_l$ and column blocks $k, k_l$ is either of the following forms
$ \begin{pmatrix}
	{\bf P}_s& {\bf X}\\
	{\bf X} & {\bf P}_s
\end{pmatrix} $or 
$\begin{pmatrix}
	{\bf X}& {\bf P}_s\\
	{\bf P}_s & {\bf X}
\end{pmatrix}.$ 
\subsection{Reduce Phase}
Receiving the sequences $\{{\bf X}_j\}_{j \in [K_1]\backslash k}$, each real node $k$ decodes all IVs of its output function, i.e., $\{v_{d_k,n} :  n \in [N])\}$
with the help of IVs $\{v_{q,n} :  q \in [K_2], W_n \in B_{f}, B_f\in \M_{\beta_k}, f \in [F]) \}$ it has access to, and finally computes the output function assigned to it.

The real node $k$ needs to obtain $\{v_{d_k,n} :  W_n \in B_{f}, B_f\notin \M_{\beta_k},f \in [F] \}$, i.e., the set of IVs required for the output function $\phi_{d_k}$ from the files not accessible to it (from the files in $B_f$ such that $ f \in [F]$ and $p_{f,{\beta_k}}\neq *$). Consider $s \in C_k$. 
Let $p_{f,\beta_k} = t \in T_{k}$. For each $k_l \in \{k_1,k_2,\ldots,k_{g_s-1}\}$ in (\ref{partition symbols}), it can compute the symbol $\U_{d_k,B_f}^{k_l}$ from the coded symbol $X_t^{k_l}$ transmitted by the real node $k_l$, i.e., 
\begin{align}
	\label{node l transmission}
	X_t^{k_l} = \bigoplus_{(\hat{f},\hat{j})\in \J_t^{k_l}} \U_{y_{\hat{j}},B_{\hat{f}}}^{{k_l}}
\end{align}
for $\J_t^{k_l} \!=\!   \{(\hat{f},\hat{j}) \!:\! p_{\hat{f},\hat{j}} \!=\!t, \hat{f}  \! \in\! [F],\hat{j} \! \in\! \{[(k_h-1)K_2\!+\!1, 
k_hK_2]:  k_h \in k \cup \{k_1,k_2,\ldots,k_{g_s-1}\}\backslash k_l\}\}$.
In (\ref{node l transmission}), for $k_h \neq k$, the $2 \times 2$ sub-block array formed by row blocks $i, i_h$ and column blocks $k, k_h$ is either of the following forms
$ \begin{pmatrix}
	{\bf P}_s& {\bf X}\\
	{\bf X} & {\bf P}_s
\end{pmatrix} $or 
$\begin{pmatrix}
	{\bf X}& {\bf P}_s\\
	{\bf P}_s & {\bf X}
\end{pmatrix}$, for integers $i$ and $i_h$ such that $s$ is present in the row indexed by $i$ and $i_h$ of PDA ${\bf P}^{(1)}$. Hence, the real node $k$ can compute $\U_{y_{\hat{j}},B_{\hat{f}}}^{k_l}$ for each $\hat{j} \in \{[(k_h-1)K_2+1,$ $k_hK_2]$ such that $p_{\hat{f},\hat{j}}=t$. If $k_h=k$, then the real node can compute $\U_{y_{\hat{j}},B_{\hat{f}}}^{k_l}$ for each $\hat{j} \in [(k-1)K_2+1,kK_2]\backslash \beta_k$ such that $p_{\hat{f},\hat{j}}=t$, since $p_{\hat{f},\beta_k}=*$ by the definition of PDA. For $\hat{j} = \beta_k$, $p_{\hat{f},\hat{j}}=p_{{f},{\beta_k}} = t$ implies $\hat{f} = f$ by condition A3-1 of Definition \ref{def pda}. Therefore, the real node $k$ can retrieve the symbol $\U_{d_k,B_f}^{k_l}$ from the coded symbol in (\ref{node l transmission}) by canceling out the rest of the symbols (since $d_k= y_{\beta_k}$). By collecting all the symbols in (\ref{partition symbols}), the real node $k$ can compute the output function $\phi_{d_k}$.

Next, we compute the computation and communication loads for this scheme. Each real node $k \in [K_1]$ stores $Z_1F_2 + (F_1-Z_1)Z_2$ batches of files. Hence, the computation load is
\begin{align}
	r&=\frac{(Z_1F_2 + (F_1-Z_1)Z_2)\eta K_1 }{\eta F_1F_2} \nonumber \\&=\frac{Z_1K_1}{F_1}+ \left (1-\frac{Z_1}{F_1} \right )\frac{Z_2K_1}{F_2}. 
\end{align}
For each $s \in [S_1]$ occurring $g_s$ times, there are $g_sS_2$ associated sequences sent, each of size $\frac{\eta \alpha}{(g_s-1)}$ bits by (\ref{transmission}). Let $S^g_1 $ denote the number of integers which appears exactly  $g$ times in the array ${\bf P}^{(1)}$. 
The communication load is given by
\begin{align}
	L_p &= \frac{1}{K_1N\alpha}\sum_{s=1}^{S_1} \frac{ g_s S_2\eta \alpha}{(g_s-1)} =  \frac{\eta }{\eta K_1F_1F_2 }\sum_{g=2}^{K_1} \frac{gS_2S_1^g}{(g-1)} 
	\nonumber\\&=  \frac{S_2}{K_1F_1F_2} \left (\sum_{g=2}^{K_1} S_1^g +\sum_{g=2}^{K_1} \frac{S_1^g}{(g-1)}  \right )\nonumber\\
	&=\frac{S_2}{K_1F_1F_2} \left (S_1 +\sum_{g=2}^{K_1} \frac{S_1^g}{ (g-1)}  \right ). 
\end{align}
\subsection{Privacy}
By our construction, using the PDAs, the allocation of file batches to each effective node remains fixed. Consequently, $({\bf X}_1, \ldots, {\bf X}_{K_1})$ depend solely on the output functions assigned to the effective nodes. Since $a_j$, for $j \in [K_2]$, is selected independently and uniformly over $[K_2]$, the variable $\beta_j$ is also independently and uniformly distributed over $[(j-1)K_2+1, $ $jK_2]$. 
For any permutation ${\bf p}$ of $[K_2]$, and for any $i \in [K_2]$, $j,k \in [K_1]$ where $j \neq k$ (with $i$ being the $p$-th element of ${\bf p}$), the following holds
\begin{align}
	& \Pr [ (y_{(j-1)K_2+1}, \ldots, y_{jK_2}) = {\bf p} \,|\, d_j = i, d_k, \mathcal{M}_k ] \nonumber \\
	&= \Pr [ (y_{(j-1)K_2+1}, \ldots, y_{jK_2}) = {\bf p} \,|\, d_j = i ] \label{eq:indep_other_info} \\
	&= \Pr [a_j = p \,|\, d_j = i ] \cdot \nonumber \\ & \quad \Pr [ (y_{(j-1)K_2+1}, \ldots, y_{p-1}, y_{p+1}, \ldots, y_{jK_2}) \,|\, a_j \!=\! p, d_j \!=\! i ] \nonumber \\
	&= \frac{1}{K_2} \cdot \nonumber \\ & \quad \Pr [ (y_{(j-1)K_2+1}, \ldots, y_{p-1}, y_{p+1}, \ldots, y_{jK_2}) \,|\, a_j \!=\! p, d_j \!=\! i ] \label{eq:Sk_uniform} \\
	&= \frac{1}{K_2} \cdot \frac{1}{(K_2-1)!} \label{eq:uniformity_other_demand}
	= \frac{1}{K_2!}.
\end{align}
where:
\begin{itemize}
	\item \eqref{eq:indep_other_info} holds because, given $d_j$, the output function assignments of effective nodes outside the range $[(j-1)K_2+1,$ $jK_2]$ are independent of the stored batches and the output function assignments of other nodes;
	\item \eqref{eq:Sk_uniform} holds because $a_j$ is uniformly distributed over $[K_2]$, independently of $d_j$; and
	\item \eqref{eq:uniformity_other_demand} follows because, given $a_j$ and $d_j$, the output functions of nodes in $[(j-1)K_2+1,jK_2]$ are uniformly selected from all permutations of $[K_2]$, with the $a_j$-th element fixed as $d_j$.
\end{itemize}
For nodes outside the column block of node $k$, i.e., nodes not in $
[(k-1)K_2 + 1, kK_2],$
their permutations ${\bf y}_j$ depend only on their own $d_j$ and $a_j$, which are independent of $(d_k, \M_k)$, where 	${\bf y}_j = (y_{(j - 1)K_2 + 1}, \ldots, y_{jK_2})$. Thus, we have
$H({\bf y}_j \mid d_k, \M_k) = H({\bf y}_j).$ From \eqref{eq:uniformity_other_demand}, we deduce that $\Pr [ {\bf y}_j \,|\, d_j, d_k, \mathcal{M}_k ]$ is independent of $(d_j, d_k, \mathcal{M}_k)$. This implies that the conditional mutual information can be expressed as follows:
\begin{align}
	\label{indep}
	I({\bf y}_j ; d_j \mid d_k, \M_k) &= H({\bf y}_j \mid d_k, \M_k) - H({\bf y}_j \mid d_j, d_k, \M_k) \nonumber \\
	&= H({\bf y}_j ) - H({\bf y}_j ) = 0.
\end{align}
The mutual information can be decomposed as
\begin{align}
		I({\bf y}_1, \ldots, {\bf y}_{K_1}; d_1, \ldots, d_{K_1} \mid d_k, \M_k) 
		= \nonumber \\   \sum_{j=1}^{K_1} I\big({\bf y}_j; d_j \mid d_k, \M_k\big),
\end{align}
because the output function assignments for nodes in different column blocks are independent given $d_k$ and $\M_k$.

For any $j \ne k$, the output function assignments ${\bf y}_j$ for the effective nodes in column block $j$ are independent of $d_j$ given $d_k$ and $\M_k$ from (\ref{indep}).
For $j = k$, since $d_k$ is already given, the mutual information is zero:
\begin{align}
	I\big({\bf y}_k; d_k \mid d_k, \M_k\big) = 0.
\end{align}
Hence, we have
\begin{align}
	I({\bf y}_1, \ldots, {\bf y}_{K_1}; d_1, \ldots, d_{K_1} \mid d_k, \M_k) 
	& = 0.
\end{align}
This shows that $\{{\bf y}_j\}_{j \in [K_1]}$ is independent of $\{d_j\}_{j \in [K_1]}$ when conditioned on $(d_k, \M_k)$. Therefore, the privacy constraint is satisfied, i.e., no node $k$ can infer any information about another node $j$'s output function index $d_j$ from the broadcast queries $\{{\bf y}_\ell\}_{\ell \in [K_1]}$, beyond what is already known from $(d_k, \M_k)$.

\section{Proof of Correctness of {\bf Construction \ref{CON1}}}
\label{proof cons1}
For each $r\in [K-1]$, the PDA \({\bf P}^{(1)}_r\) constructed using \textbf{Algorithm \ref{algo2}} with \(t = r\) is a $(r+1)$-$\left ( K, {K \choose r},{K-1 \choose r-1}, {K \choose r+1} \right )$ PDA.

Each array \({\bf P} \in \mathcal{P}\) is structured as a block array consisting of \({K \choose r}\) row blocks indexed by \(\left [{K \choose r} \right ]\) and \(K\) column blocks indexed by \([K]\). The block entries are denoted as \(\{{\bf p}_{f,k} : f \in \left [{K \choose r} \right ], k \in [K]\}\). These blocks include the arrays \({\bf X}\) and \({\bf P}_s = [(s-1)Q+1,  sQ]\) for \(s \in \left [{K \choose r+1} \right ]\), each of size \(1 \times Q\). Consequently, the overall size of \({\bf P}\) is \({K \choose r}  \times KQ\). 

Each column in the PDA \({\bf P}^{(1)}_r\) contains exactly ${K-1 \choose r-1}$ stars, and each star is replaced by the array \({\bf X}\), which itself contains one star per column. Additionally, each column in \({\bf P}^{(1)}_r\) contains \({K \choose r}-{K-1 \choose r-1}\) integers, with each integer \(s\) being replaced by the array \({\bf P}_s\), containing all integers in its defined range. Thus, each column in \({\bf P}\) contains ${K-1 \choose r-1}$ stars, satisfying condition \(A1\) of Definition \ref{def:PDA} with \(Z = {K-1 \choose r-1}\). 

The integers in \({\bf P}_s\) lie within the range \([(s-1)Q+1, sQ]\) for each \(s \in \left[{K \choose r+1} \right]\), and the total number of integers in \({\bf P}\) is \(Q{K \choose r+1}\), spanning the range \(\left[Q{K \choose r+1}\right]\). This satisfies condition \(A2\) of Definition \ref{def:PDA}.

From \textbf{Construction \ref{CON1}}, for two distinct block entries \({\bf p}_{f_1,k_1}\) and \({\bf p}_{f_2,k_2}\), \({\bf p}_{f_1,k_1} = {\bf p}_{f_2,k_2} = {\bf P}_s\) for some \(s \in \left[{K \choose r+1}\right]\) only if \(f_1 \neq f_2\) and \(k_1 \neq k_2\). These entries belong to distinct row and column blocks. Furthermore, \({\bf p}_{f_1,k_2} = {\bf p}_{f_2,k_1} = {\bf X}\), meaning the corresponding \(2 \times 2\) sub-block array formed by row blocks \(f_1, f_2\) and column blocks \(k_1, k_2\) must have one of the following forms:
\[
\begin{pmatrix}
	{\bf P}_s & {\bf X} \\
	{\bf X} & {\bf P}_s
\end{pmatrix}
\quad \text{or} \quad
\begin{pmatrix}
	{\bf X} & {\bf P}_s \\
	{\bf P}_s & {\bf X}
\end{pmatrix}.
\]

Finally, for any two distinct values \(s_1, s_2 \in \left[{K \choose r+1}\right]\), the integers in \({\bf P}_{s_1}\) and \({\bf P}_{s_2}\) are disjoint. Thus, each entry in \({\bf P}_s\) satisfies condition \(A3\) of Definition \ref{def:PDA}. 

In conclusion, the array \({\bf P}\) is a $\left ( KQ, {K \choose r},{K-1 \choose r-1}, Q{K \choose r+1} \right )$ PDA.
\section{Proof of Theorem \ref{thm2}}
\label{proof thm2}
For $r \in [K-1]$, the input database is split into ${K \choose r}$ disjoint batches. 
The  $\left ( KQ, {K \choose r},{K-1 \choose r-1}, Q{K \choose r+1} \right )$ PDA in $\mathcal{P}$ corresponds to a DC model with $K$ nodes, with the rows corresponding to the batches and the column corresponding to the effective nodes.

The computing phases follow from the proof of Theorem \ref{thm1}. Since the number of stars in each column of the PDA is ${K-1 \choose r-1}$, the computation load is given by
\begin{align}
	\frac{{K-1 \choose r-1}K}{{K \choose r}} =r.
\end{align}
For each $s \in \left[{K \choose r+1}\right]$ occurring ${r+1}$ times, there are $(r+1)Q$ associated sequences sent, each of size $\frac{\eta \alpha}{r}$ bits. Hence, the  communication load is given by
\begin{align}
	L_p &= \frac{1}{KN\alpha}\sum_{s=1}^{{K \choose r+1}} (r+1) Q\frac{\eta \alpha}{r} \nonumber\\&= \frac{{K \choose r+1}}{{K \choose r}} \cdot \frac{\eta (r+1)Q}{\eta Kr } \nonumber\\&= \frac{(K - r)(r+1)Q}{{(r+1)K  r}} 
	\nonumber\\&=  \frac{Q}{r} \left ( 1- \frac{r}{K}\right). 
\end{align}
\section{Proof of Correctness of {\bf Construction \ref{CON2}}}
\label{proof cons2}
The set of PDAs in \(\mathcal{P}^{(1)}\) is generated using \textbf{Algorithm \ref{algo2}} with \(t = r_1\) and $D=K$, resulting in a $(r_1+1)$-\(\left( K, {K \choose r_1}, {K-1 \choose r_1-1}, {K \choose r_1+1} \right)\) PDA for \(r_1 \in [K-1]\). Similarly, the PDAs in \(\mathcal{P}^{(2)}\) are constructed using \textbf{Algorithm \ref{algo2}} with \(t = r_2\) and $D=Q$, forming a $(r_2+1)$-\(\left( Q, {Q \choose r_2}, {Q-1 \choose r_2-1}, {Q \choose r_2+1} \right)\) PDA, where \(r_2 \in [Q-1]\).  

For each \(r_1 \in [K-1]\) and \(r_2 \in [Q-1]\), we select a PDA \(\mathbf{P}^{(1)}\) from \(\mathcal{P}^{(1)}\) and a PDA \(\mathbf{P}^{(2)}\) from \(\mathcal{P}^{(2)}\). Using these, we construct an extended PDA \(\mathbf{P}\) with parameters  
$
\left( KQ, {K \choose r_1}{Q \choose r_2}, {K-1 \choose r_1-1}{Q \choose r_2+1} \left(\frac{Q}{r_2} - \frac{K-r_1}{r_1} \right), {K \choose r_1+1} {Q \choose r_2+1} \right)
$
following \textbf{Algorithm \ref{algo1}}. The collection of PDAs in \(\mathcal{P}\) consists of all such PDAs obtained by varying \(r_1\) from \(1\) to \(K-1\) and \(r_2\) from \(1\) to \(Q-1\).
\section{Proof of Theorem \ref{thm3}}
\label{proof thm3}
For each $r_1 \in [K-1]$ and $r_2=[Q-1]$, the input database is split into $F={K \choose r_1}{Q \choose r_2}$ disjoint batches. 
The extended PDA corresponds to a DC model, with the rows corresponding to the batches and the column corresponding to the effective nodes. 
The computing phases follow from the proof of Theorem \ref{thm1}. Hence, the computation load is given by
\begin{align}
	r=&\frac{Z_1K_1}{F_1}+ \left (1-\frac{Z_1}{F_1} \right )\frac{Z_2K_1}{F_2}\nonumber\\
	=&\frac{{K-1 \choose r_1-1}K}{{K \choose r_1}}+\left( 1-\frac{{K-1 \choose r_1-1}}{{K \choose r_1}}  \right) \frac{{Q-1 \choose r_2-1}K}{{Q \choose r_2}}\nonumber \\
	=&\frac{(K-1)! K (K-r_1)!r_1!}{(r_1-1)!(K-r)! K! } \nonumber \\
	&+\left( 1-\frac{(K-1)!  (K-r)!r_1!}{(r_1-1)!(K-r_1)! K! } \right) \frac{(Q-1)! K (Q-r)!r_2!}{(r_2-1)!(Q-r_2)! Q! }\nonumber \\
	=&r_1+\left( 1-\frac{r_1}{K}  \right) \frac{r_2K}{Q} \nonumber\\
	=&r_1 + \frac{(K-r_1)r_2}{Q}.
\end{align}
The communication load is given by 
$      \frac{S_2}{K_1F_1F_2} \left (S_1 +\sum_{g=2}^{K_1} \frac{S_1^g}{ (g-1)}  \right )$
where $S_1^g$ is the number of integers in $[S_1]$ which appear exactly $g$ times in the PDA ${\bf P}^{(1)}$.  In this case all the integers appear exactly ${r_1+1}$ times. Hence, we have 
\begin{align}
	L_{p} =& \frac{ {Q \choose r_2+1}}{K{K \choose r_1}{Q \choose r_2} } \left ({K \choose r_1+1} +\frac{{K \choose r_1+1}}{ r_1}  \right )\nonumber\\
	=& \frac{ Q! r_1! (K-r_1)! r_2! (Q-r_2)!}{K(r_2+1)! (Q-(r_2+1))!K!  Q!   } \nonumber\\
	& \left (\frac{K!  }{(r_1+1)! (K-(r_1+1))!}  \right ) \left (\frac{r_1+1  }{r_1 }\right)\nonumber\\
	=& \frac{ {Q - r_2}}{K{(r_2+1)} }\left (\frac{ K-r_1}{ r_1}  \right )\nonumber\\
	=&\frac{1}{r_1} \left(1 - \frac{r_1}{K}\right)\left(\frac{Q-r_2}{r_2+1}\right).
\end{align}
\bibliographystyle{IEEEtran}
\begin{IEEEbiography}[{\includegraphics[width=1in,height=1.25in,keepaspectratio]{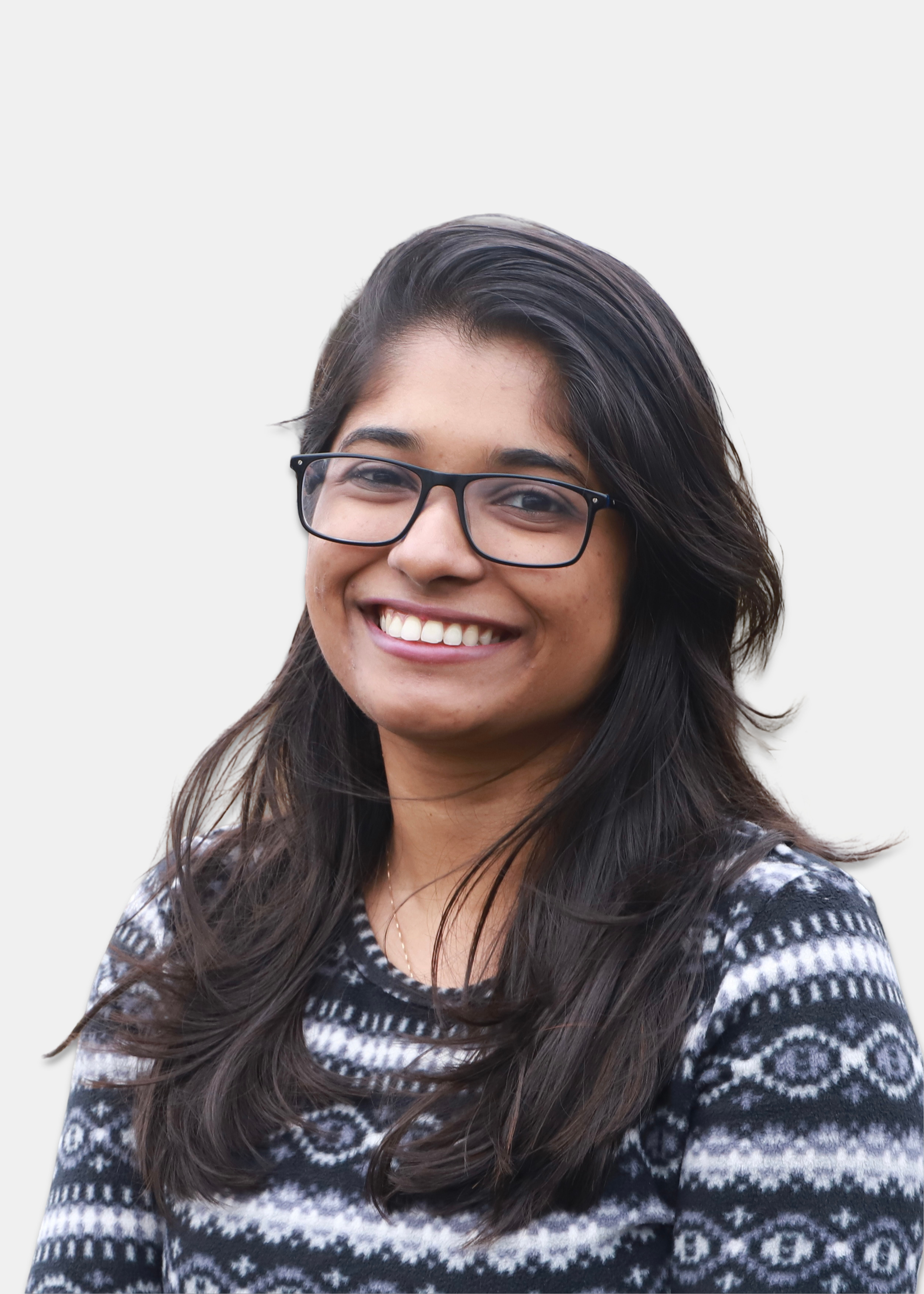}}] {Shanuja Sasi} received her B.Tech degree in Electronics and Communication Engineering from the National Institute of Technology, Calicut, in 2016. She earned her M.Sc. and Ph.D. degrees from the Department of Electrical Communication Engineering at the Indian Institute of Science (IISc), Bangalore, in 2021. In 2019, she was awarded the Overseas Visiting Doctoral Fellowship (OVDF) by the Science and Engineering Research Board (SERB), during which she spent a year as a visiting scholar at Purdue University, Indiana.
	
	From 2021 to 2023, she worked as a Senior Engineer at Qualcomm, Bangalore, where she contributed to advanced communication systems research. Between 2023 and 2025, she served as a Postdoctoral Researcher at Linköping University, Sweden. In 2024, she received the INSPIRE Faculty Fellowship from the Department of Science and Technology (DST), Government of India, and currently holds the position of INSPIRE Faculty Fellow at the Indian Institute of Technology, Kanpur.
	
	Her research interests lie in the areas of privacy and security in distributed computing, with a particular focus on coded caching, gradient coding, coding theory, and their applications in machine learning and federated learning.
\end{IEEEbiography}
\begin{IEEEbiography}[{\includegraphics[width=1in,height=1.25in,keepaspectratio]{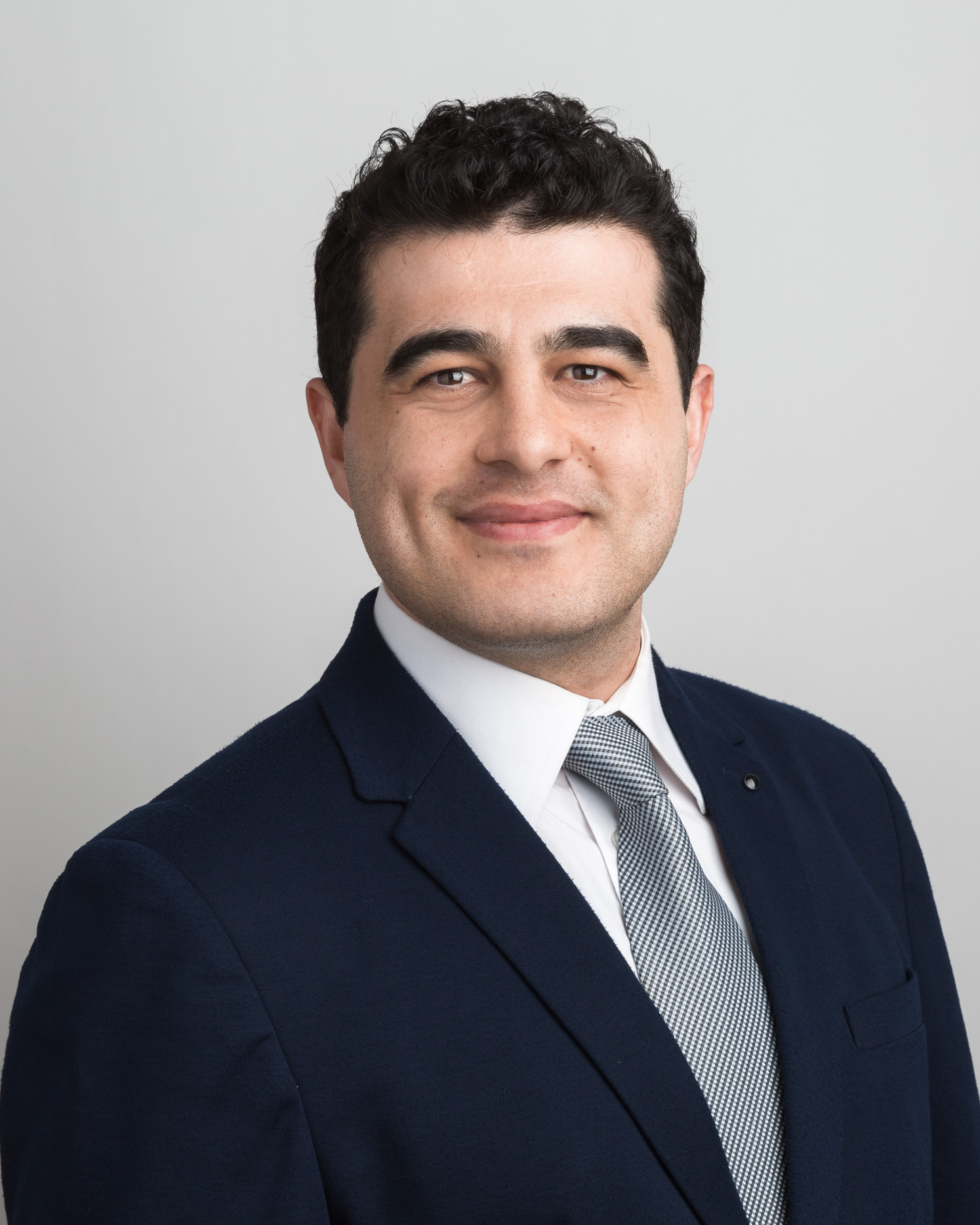}}] {Onur Günlü}
	(Senior Member, IEEE) received the B.Sc. degree (Highest Distinction) in Electrical and Electronics Engineering from Bilkent University, Turkey in 2011; M.Sc. (Highest Distinction) and Dr.-Ing. (Ph.D. equivalent) degrees in Communications Engineering both from the Technical University of Munich (TUM), Germany in 2013 and 2018, respectively. He was a Working Student in the Communication Systems division of Intel Mobile Communications (IMC), now Apple Inc., in Munich, Germany during November 2012 - March 2013. Onur worked as a Research and Teaching Assistant at TUM Chair of Communications Engineering (LNT) between February 2014 - May 2019. As a Visiting Researcher, among more than twenty Research Stays at Top Universities and Companies, he was at TU Eindhoven, Netherlands during February 2018 - March 2018. Onur was a Visiting Research Group Leader at Georgia Institute of Technology, Atlanta, USA during February 2022 - March 2022. He was also a Visiting Professor at TU Dresden, Germany during February 2023 - March 2023. Following Research Associate and Group Leader positions at TUM, TU Berlin, and the University of Siegen, he joined Linköping University in October 2022 as an ELLIIT Assistant Professor and obtained tenure as an Associate Professor leading the Information Theory and Security Laboratory (ITSL) in August 2024. He obtained a Swedish Docent (Habilitation) of Information Theory title in December 2023 and became an IEEE Senior Member in July 2024. Since September 2025, Onur has been a Full Professor and Institute Head at the Lehrstuhl für Nachrichtentechnik at TU Dortmund and a Guest Professor at Link{\"o}ping University. 
	
	He has received the 2025 IEEE Information Theory Society - Joy Thomas Tutorial Paper Award, the 2023 ZENITH Research and Career Development Award, 2021 IEEE Transactions on Communications - Exemplary Reviewer Award, and the prestigious VDE Information Technology Society (ITG) 2021 Johann-Philipp-Reis Award. His research interests include distributed function computation, information-theoretic privacy and security, coding theory, integrated sensing and communication, and private learning. Among his publications is the book \emph{Key Agreement with Physical Unclonable Functions and Biometric Identifiers} (Dr. Hut Verlag, 2019). He serves as an Associate Editor for \textsc{IEEE Journal on Selected Areas in Communications}, \textsc{IEEE Transactions on Communications}, and \textsc{Entropy}, and recently was an Associate Editor for \textsc{EURASIP Journal on Wireless Communications and Networking} and a Guest Editor for \textsc{IEEE Journal on Selected Areas in Information Theory}. He also serves as a Board Member and Secretary of the IEEE Sweden VT/COM/IT Joint Chapter and as a Working Group Leader for EU COST Action 6G Physical Layer Security (6G-PHYSEC).
\end{IEEEbiography}
\end{document}